\documentclass{article}

\usepackage[final,nonatbib]{neurips_2024}

\usepackage[utf8]{inputenc} 
\usepackage[T1]{fontenc}    
\usepackage{hyperref}       
\usepackage{url}            
\usepackage{booktabs}       
\usepackage{amsfonts}       
\usepackage{nicefrac}       
\usepackage{microtype}      
\usepackage{xcolor}         
\usepackage{multirow}
\usepackage{booktabs}
\usepackage{diagbox}
\usepackage{enumitem}
\usepackage{threeparttable}
\usepackage{xspace}
\usepackage{adjustbox}
\usepackage{amsmath}
\usepackage{wrapfig}

\newcommand{\modelname}{DeltaDock\xspace}
\newcommand{\sitemodel}{CPLA\xspace}
\newcommand{\refinemodel}{Bi-EGMN\xspace}

\title{\modelname: A Unified Framework for Accurate, Efficient, and Physically Reliable Molecular Docking}

\author{Jiaxian Yan$^1$, Zaixi Zhang$^1$, Jintao Zhu$^2$, Kai Zhang$^1$, Jianfeng Pei$^2$, Qi Liu$^{1}$\thanks{Qi Liu is the corresponding author.}\\
$^1$State Key Laboratory of Cognitive Intelligence, University of Science and Technology of China\\
$^2$Center for Quantitative Biology, \\ Academy for Advanced Interdisciplinary Studies, Peking University \\
\texttt{ \{jiaxianyan, zaixi, sa517494\}@mail.ustc.edu.cn, zhujt@stu.pku.edu.cn,} \\
\texttt{jfpei@pku.edu.cn, qiliuql@ustc.edu.cn}
}

\begin{document}

\maketitle
\vspace{-0.3in}
\begin{abstract}
Molecular docking, a technique for predicting ligand binding poses, is crucial in structure-based drug design for understanding protein-ligand interactions. Recent advancements in docking methods, particularly those leveraging geometric deep learning (GDL), have demonstrated significant efficiency and accuracy advantages over traditional sampling methods. Despite these advancements, current methods are often tailored for specific docking settings, and limitations such as the neglect of protein side-chain structures, difficulties in handling large binding pockets, and challenges in predicting physically valid structures exist. To accommodate various docking settings and achieve accurate, efficient, and physically reliable docking, we propose a novel two-stage docking framework, \modelname, consisting of pocket prediction and site-specific docking. We innovatively reframe the pocket prediction task as a pocket-ligand alignment problem rather than direct prediction in the first stage. Then we follow a bi-level coarse-to-fine iterative refinement process to perform site-specific docking. Comprehensive experiments demonstrate the superior performance of \modelname. Notably, in the blind docking setting, \modelname achieves a 31\% relative improvement over the docking success rate compared with the previous state-of-the-art GDL model. With the consideration of physical validity, this improvement increases to about 300\%.\footnote[2]{All codes and data will be released on \url{https://github.com/jiaxianyan/DeltaDock}.}
\end{abstract}
\vspace{-0.1in}
\section{Introduction}
\label{sec:intro}
Recent advancement in geometric deep learning~(GDL)~\cite{Bronstein2021GeometricDL, zhang2023fullatom, Zhang2023LearningSP} presents an innovative and promising molecular docking paradigm to predict and understand the interactions between target proteins and drugs, which is of paramount importance for drug discovery~\cite{Du2016InsightsIP, Li2021StructureawareIG}. 
Unlike traditional docking methods that employ optimization algorithms to sample and identify best binding poses~\cite{Lyu2019UltralargeLD, Bender2021APG}, GDL methods interpret molecular docking as either a regression or generation task, eliminating the need for intensive candidate sampling~\cite{Strk2022EquiBindGD, Corso2022DiffDockDS, Zhang2023ASS}.
Studies have demonstrated that GDL methods outperform their traditional counterparts, delivering enhancements in both the accuracy of binding pose predictions, as measured by the root-mean-square deviation (RMSD) metric, and the inference efficiency~\cite{Pei2023FABindFA, Zhang2022E3BindAE}.

According to whether a prior pocket is given, molecular docking can be divided into blind and site-specific docking~\cite{Hassan2017ProteinLigandBD}. Traditional sampling methods adeptly navigate both scenarios, primarily differing in the scope of the search space they explore. In contrast, GDL methods typically specialize in either one. For instance, EquiBind~\cite{Strk2022EquiBindGD}, and DiffDock~\cite{Corso2022DiffDockDS} are designed for blind docking, neglecting the incorporation of binding pockets. Uni-Mol~\cite{Zhou2023UniMolAU} and DiffBind-FR~\cite{Zhu2023DiffBindFRAS} concentrate on site-specific docking and only protein atomic level structure within a defined radius (usually 6-12 \AA ) of the co-crystal is modeled.
Despite some progress, these methods not only fail to handle two docking settings smoothly like traditional methods, but also confronted with certain limitations. For blind docking methods, they ignore the fine-grained protein side-chain structure. Regarding the site-specific docking methods, when dealing with pockets larger than the predetermined cutoff or when there is a requirement to model extensive pocket surrounding structures to account for long-range interactions, these methods significantly deteriorate in performance~\cite{Buttenschoen2023PoseBustersAD} and the demand for computational resources can escalate significantly, as evidenced in Appendix.\ref{apdsec:data_statistics} and Appendix.\ref{apdsec:large_pocket_exmaple}. 

Besides these challenges, current GDL methods face additional limitations due to the lack of inductive biases, such as penalties for steric clashes or constraints on ligand mobility, leading to the generation of unrealistic docking poses. Buttenschoen et al.~\cite{Buttenschoen2023PoseBustersAD} proposed the PoseBusters test suit to verify and highlight these problems. In addition to the RMSD between predicted and ground-truth poses, the test suite incorporates 18 checks, encompassing chemical validity and consistency, intramolecular validity, and intermolecular validity.
According to the test suite, the previously highest-performing method, DiffDock, achieves a success rate of only 14\%. This is significantly lower than the 38\% success rate achieved when chemical validity is not taken into account.

To resolve these problems, we propose \textbf{\modelname}, a unified GDL framework for accurate, efficient, and physically valid docking. \modelname is a two-stage framework consisting of a pocket prediction stage and a site-specific docking stage.
With \textbf{"Delta"}, we mean that the optimal poses are predicted by iteratively refining the input structures in the second docking stage.
The first pocket prediction stage is specialized for blind docking, where a binding pocket is identified from a set of candidates through a novel contrastive pocket-ligand alignment module \sitemodel.
Then in the second stage, within the pockets predefined or selected by \sitemodel, binding structures are predicted in a bi-level coarse-to-fine iterative refinement module \refinemodel.
This module prioritizes the residue-level structure covered by a large outer box (Fig.\ref{apdfig:bilevel_example}) for pose positioning and coarse structure prediction. And the atom-level structure, within a relatively small radius from the coarse structure, is characterized for more refined predictions.
In particular, the module incorporates (i) a GPU-accelerated pose sampling algorithm generating high-quality initial structure, (ii) a training objective imposing penalties for steric clashes and constraints on ligand mobility, and (iii) a rapid post-processing step composing torsional alignment and energy minimization for structure correction.

To accommodate two different docking settings, \modelname is specially designed as a two-stage framework rather than an end-to-end framework. 
Particularly, the pocket-ligand alignment module is inspired by the observation shown in Fig.\ref{fig:apd_dcc_total}. Existing pocket prediction methods generally achieve a recall rate of just 70\%-80\%. However, when combining all possible pockets predicted by multiple methods, this recall rate reaches nearly 95\%. According to this result, we shift the focus from designing increasingly powerful pocket prediction models to developing strategies for the effective selection of a candidate pocket from an ensemble of predicted pockets. The pocket prediction task is thus reframed as a pocket-ligand alignment problem innovatively.
Regarding the site-specific docking stage, the key idea is to accurately predict reliable poses.
Based on the proposed bi-level iterative refinement model, several components presented above are introduced additionally.
Among them, the pose sampling algorithm is adopted for structure initialization, as previous works on structure prediction~\cite{ Reidenbach2023CoarsenConfEC} have demonstrated the importance of a good initial structure.
Other two components, namely the physics-informed training object and the fast structure correction step, are leveraged to ensure physical validity.

To demonstrate the effectiveness of \modelname, we performed comprehensive experiments to evaluate its predictive accuracy, efficiency, generalizability, and ability to predict physically valid binding poses. The experimental outcomes indicate that \modelname consistently surpasses the baseline methods in both blind docking and site-specific docking settings while maintaining remarkable computational efficiency.
Notably, in the blind docking setting, \modelname exceeded the performance of the previous SOTA GDL method, DiffDock, by 30.8\% in terms of the docking success rate, and it required only approximately 3.0 seconds per protein-ligand pair.
With the consideration of physical validity, this improvement increases to approximately 300\% on the PoseBusters benchmark.

\vspace{-0.1in}
\section{Related Work}
\label{sec:related}
\subsection{Sampling-based Docking}
Traditional docking methods, epitomized by the likes of VINA~\cite{Trott2010AutoDockVI} and SMINA~\cite{Koes2013LessonsLI}, operate on a "sampling-and-scoring" paradigm to identify the best binding pose. 
Optimization algorithms such as BFGS~\cite{Nocedal2000NumericalO} are used to sample optimal poses within the defined search space on CPUs. This process, which involves a significant number of steps and multiple copies, is rather computationally intensive.
Recent studies have attempted to speed up the sampling process using GPUs. Notable examples are Vina-GPU~\cite{Ding2023VinaGPU2F}, Uni-Dock~\cite{Yu2023UniDockGD}, and DSDP~\cite{Huang2023DSDPAB}, which use more copies and shorter search steps to fully leverage the parallel computational power of GPUs. This approach has demonstrated substantial efficacy, achieving a speed increase of an order of magnitude compared to prior CPU-based methods.

\subsection{Geometric Deep Learning-based Docking}
GDL introduces a new paradigm in molecular docking, where the sampling process is bypassed by interpreting molecular docking as either a regression task or a generation task~\cite{Strk2022EquiBindGD,Corso2022DiffDockDS}.
However, recent researches have highlighted limitations of current GDL methods, such as neglect of protein side-chain structures~\cite{Zhu2023DiffBindFRAS}, difficulties in handling large binding pockets, and challenges in predicting physically valid structures~\cite{Buttenschoen2023PoseBustersAD}. 
Compared with physically reliable sampling-based methods, especially recent developed GPU-accelerated methods, the existing limitations hinder the practical application of GDL methods.
To address these concerns, in this work, we propose \modelname to overcome these problems and accomplish efficient, accurate, and physical reliable docking.

\subsection{Binding Pocket Prediction}
As the foundation of structure-based drug design, binding pocket prediction has attracted expansive attention. A variety of methods have been developed for this task, encompassing traditional computational methods, such as Fpocket~\cite{Guilloux2009FpocketAO}, machine learning~(ML) methods, such as P2Rank~\cite{Krivk2018P2RankML}, and GDL methods, such as PUResNet~\cite{Kandel2021PUResNetPO}.
These methods generally adopt ligand-free approaches and focus on predicting all potential binding sites within individual proteins. Recent blind docking methods, DSDP and FABind, apply pocket prediction for target ligands to reduce the docking search space, which is of great help to fast and accurate blind docking.
In this study, our proposed model, \modelname, also prioritizes defining a pocket for blind docking. However, instead of improving model architecture for pocket prediction like previous methods, \modelname reframe the pocket prediction task as a pocket-ligand alignment problem and employ contrastive learning to select a candidate pocket from the combined pockets set.

\vspace{-0.1in}
 
\begin{figure}[t]
  \centering
  \includegraphics[width=\linewidth]{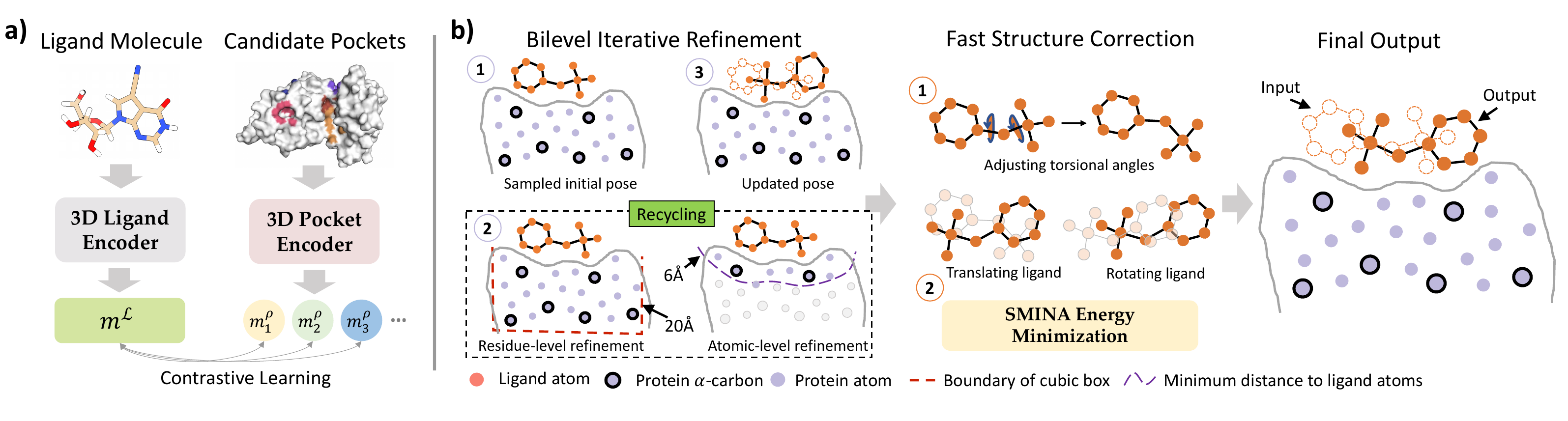}
  \caption{
  The overview of \modelname's two modules.
  (a) The pocket-ligand alignment module \sitemodel. Contrastive learning is adopted to maximize the correspondence between target pocket and ligand embeddings for training. During inference, the pocket with the highest similarity of the ligand is selected.
  (b) The bi-level iterative refinement module \refinemodel. Initialized with a high-quality sampled pose, the module first performs a coarse-to-fine iterative refinement. This process generates progressively refined ligand poses utilizing a recycling strategy. To guarantee the physical plausibility of the predicted poses, a two-step fast structure correction is subsequently applied. This correction involves torsion angle alignment followed by energy minimization based on the SMINA.
  }
  \label{fig:framework}
 \end{figure}

 \section{\modelname Framework}
\label{sec:methods}
\subsection{Preliminaries}
\noindent\textbf{Notations.} In this work, the separate structures of a protein $\mathcal{P}$ and a ligand $\mathcal{L}$ are used as inputs (Fig.~\ref{fig:framework}). Both molecules are initially encoded as graphs, and we denote a molecule graph as $\mathcal{G}=(\mathcal{V}, \mathcal{E})$, where $\mathcal{V}$ and $\mathcal{E}$ represent the node set and edge set respectively. Each node $v_i \in \mathcal{V}$ is associated with a coordinate $x_i$ and a feature vector $h_i$. Each edge $(i,j)\in \mathcal{E}$ is associated with an edge feature vector $e_{ij}$. 
For the ligand $\mathcal{L}$ and ligand graph $\mathcal{G}^\mathcal{L}$, $v^\mathcal{L}_i$ represents the $i$-th atom in the ligand and $x^\mathcal{L}_i$ corresponds to the atom's coordinate. 
For the protein $\mathcal{P}$, the situation is more complex, and two graphs based on the two structural levels of the protein are constructed. One is the protein atomic graph $\mathcal{G}^{\mathcal{P}}$, and the other is the protein residue graph $\mathcal{G}^{\mathcal{P}*}$. 
$\mathcal{G}^{\mathcal{P}}$ contains protein atomic-level information similar to ligand graph $\mathcal{G}^\mathcal{L}$, while $\mathcal{G}^{\mathcal{P}*}$ contains protein residue-level information and overlooks the side-chain structure information.
In $\mathcal{G}^{\mathcal{P}*}$, $v^{\mathcal{P}*}_i$ represents the $i$-th residue in the protein and $x^{\mathcal{P}*}_i$ corresponds to the $C_\alpha$ coordinate of this residue. 
Details of the graph construction can be found in Appendix.\ref{apd:graph_construct}.

\noindent\textbf{Overview.} Our goal is to train a model $f$ that excels in both site-specific docking and blind docking scenarios of rigid molecular docking, wherein the protein structure is fixed and only the ligand's flexibility is considered.

\textit{As depicted in Fig.~\ref{fig:framework}, \modelname comprises two modules: a pocket-ligand alignment module \textbf{\sitemodel} responsible for selecting binding pocket from a pocket candidate set, and a bi-level iterative refinement module \textbf{\refinemodel} dedicated to executing site-specific docking given the binding pockets. This design allows \modelname to handle both blind docking and site-specific docking seamlessly. In the subsequent part of this section, we will elaborate on these two modules respectively.}

\subsection{Contrastive Pocket-ligand Alignment}
\label{sec:pocket_align} 
\sitemodel treats the pocket prediction task as a pocket-ligand alignment problem. We employ a list of well-established ligand-free pocket prediction methods to generate candidate pocket sets, and then map these pockets and the target ligand into the same embedding space. The correct pocket embedding is expected to have higher similarity with ligand embedding than other pockets.
\subsubsection{Data Preprocessing}
The initial step of this module involves using RDKit~\cite{greg_landrum_2022_6798971} to generate a 3D conformer of the input ligand, as depicted in Fig.~\ref{fig:framework}.
Binding site prediction models including P2Rank and DSDP are adopted to extract druggable binding sites, and the binding sites predicted by these different methods are combined to form a set of candidate binding sites, denoted as $S = \{\varsigma_1, \varsigma_2, ...\}$, where $\varsigma_i$ represents the geometric center of $i$-th binding site. For \sitemodel, the protein pocket $\rho_i$ is defined as the residues within 15.0 \AA \; to $\varsigma_i$. 

\subsubsection{Ligand and Pocket Encoders}
To map the ligand and pockets into the embedding space, the ligand encoder Attentive-FP~(AFP)~\cite{Xiong2020PushingTB} and protein encoder Geometric Vector Perceptron~(GVP)~\cite{Jing2020LearningFP} are employed.
These encoders first extract informative ligand node and protein node representations, and the feature extraction process can be formally expressed as:
\begin{align}
            H^\mathcal{L} = AFP(\mathcal{G}^\mathcal{L}), \;
            H^{\mathcal{P}*} = GVP(\mathcal{G}^{\mathcal{P}*}),
\end{align}
where $H^\mathcal{L}$ is the ligand embedding matrix of shape $|\mathcal{V}^\mathcal{L}| \times d$ and $H^{\mathcal{P}*}$ is the protein residue embedding matrix of shape $|\mathcal{V}^{\mathcal{P}*}| \times d$. 
The ligand representations $m^\mathcal{L}$ and pocket representations $m^{\rho}_i$ are then obtained by pooling ligand nodes embedding and pocket nodes embedding:
\begin{align}
    m^\mathcal{L} = Sum(H^\mathcal{L}, \mathcal{V}^\mathcal{L}), \;
    m^{\rho}_i = Sum(H^{\mathcal{P}*}, \mathcal{V}^{\rho}_i),
\end{align}
where $\mathcal{V}^{\rho}_i$ is the protein node set of $i$-th pocket $\rho$, and the pooling operation is sum pooling. For the pocket encoder, we input the entire protein residue graph $\mathcal{G}^{\mathcal{P}*}$ rather than just the protein pocket residue graph, to incorporate global protein information into the pocket representation.

\subsubsection{Contrastive Embdding Alignment}
With ligand representation $m^\mathcal{L}$ and pocket representation $m^{\rho}_i$ in hand, we calculate the cosine similarity score:
\begin{equation}
    s_i = \frac{m^\mathcal{L} \cdot m^{\rho}_i}{\left\| m^\mathcal{L} \right\|_2 \cdot \left\| m^{\rho}_i \right\|_2}.
\end{equation}
For the candidate pockets $S = \{\varsigma_1, \varsigma_2, ...\}$, the similarity score $s_{+}$ between the target pocket and the ligand is expected to be higher than others. Thus,  we propose the contrastive learning objective:
\begin{equation}
    L = -\frac{1}{N} \cdot log \frac{exp(s_{+} / \tau)}{\sum_i exp(s_i / \tau)}, 
    \label{eq:pocketalign_loss}
\end{equation}
where $\tau$ is the temperature paramter. For blind docking, the pocket with the highest similarity score with the ligand is selected for the next docking step.

\subsection{Bi-level Iterative Refinement}
\label{sec:iterative_refine}
With a binding site $\varsigma$ predefined by the user or selected by \sitemodel, we design the bi-level iterative refinement module \refinemodel to predict binding pose within this pocket~(Fig.~\ref{fig:framework}).

\subsubsection{Inital Structure Sampling}
For an iterative refinement module, an initial structure is needed as a starting point.
Previous work on molecular 3D conformer generation~\cite{Reidenbach2023CoarsenConfEC} demonstrates the importance of a good initial structure.
Therefore, \refinemodel adopts a rapid GPU-accelerated sampling method proposed by Huang~et~al.~\cite{Huang2023DSDPAB} to sample a high-quality initial  $\mathcal{X}^{\mathcal{L}}$.
In this work, the search steps number and the search copy number are set to 40 and 384, respectively. 
Details about the search box setting can be found in the Appendix.\ref{apdsec:cpla_exp_settings}.

\subsubsection{Structure Refinement}
With input initial structure $\mathcal{X}^{\mathcal{L}}$, we iteratively update it to improve its accuracy.
As discussed in Sec.\ref{sec:intro}, the modeling of an entire binding pocket structure is crucial for the success of the process. 
Current methods either ignore the atom-level structure or model the full-atom pocket structure directly. The latter approach can significantly elevate the computational resource demand, particularly when dealing with large pockets.
To overcome these challenges and maintain high docking accuracy and efficiency, we propose a bi-level strategy in this work. In the following sections, we first present the details of the bi-level strategy. Subsequently, we discuss the \refinemodel layer, which is used to perform refinement, as depicted in Fig.~\ref{fig:framework}.

\textbf{Bi-level strategy.}
The first refinement level is the residue level, where the protein residues within a 40.0 \AA \; cubic region centered at the geometric centers of ligands are considered as pocket $\rho$. Previous work demonstrates such a range is large enough to cover the binding pocket~\cite{Lu2022TANKBindTN}.
In this context, as the full-atom structure of proteins is not considered,  the pocket residue graph $\mathcal{G}^{\mathcal{\rho}*}$ is adapted. 
The second level is the atomic level, where we set the ligand structures refined through $T$ rounds of residue level refinement as the reference structure. In this level, protein atoms within a $6.0$ \AA \; radius of the ligand atoms are considered to construct pocket atomic graph $\mathcal{G}^{\mathcal{\rho}}$ for modeling the fine-grained interaction.
The ligand coordinates $X^{a, \mathcal{L}}$ output by the last layer of atomic level refinement correspond to the final predicted structure $\hat{\mathcal{X}}^{\mathcal{L}}$.

\textbf{\refinemodel Layer}. 
The bi-level E(3)-equivariant graph matching network~(\refinemodel) layer is the model designed to calculate the protein-ligand interaction and refine the structures.
More specifically, this layer adheres to the message-passing paradigm~\cite{Gilmer2017NeuralMP} and consists of four functions: intra-message function, inter-message function, aggregate function, and update function. 

The intra-message function works to extract messages $m_{i,j}$ and $\hat{m}_{i,j}$ between a node $i$ and its neighbor nodes $j$ from the same molecule graph. $m_{i,j}$ is later used for the updating of node features and $\hat{m}_{i,j}$ for the updating node coordinates.
$\forall (i,j) \in \mathcal{E}_\mathcal{P} \cup \mathcal{E}_\mathcal{L}$, this function can be formally written as :
 \begin{align}
    d_{i,j}^{(l)} = ||x_i^{(l)} - x_j^{(l)}||, \;
    m_{i,j} = \varphi_m(h_i^{(l)}, h_j^{(l)}, d_{i,j}^{(l)},), \;
    \hat{m}_{i,j} = (x_i^{(l)} - x_j^{(l)}) \cdot \varphi_{\hat{m}}(m_{i,j}), 
 \end{align}
where $d_{i,j}^{(l)}$ is the relative distance between node $i$ and node $j$, and $\varphi$ is a MLP. 

The inter-message function works to extract messages $\mu_{i,j}$ and $\hat{\mu}_{i,j}$ between a node $i$ and its neighbor nodes $j$ from the other molecule graphs. Formally, $\forall i \in \mathcal{V}_\mathcal{P}, j \in \mathcal{V}_\mathcal{L} \, or \ i \in \mathcal{V}_\mathcal{L}, j \in \mathcal{V}_\mathcal{P}$:
 \begin{align}
    \mu_{i,j} = \varphi_\mu(h_i^{(l)}, h_j^{(l)}, d_{i,j}^{(l)}), \; 
    \hat{\mu}_{i,j} =
    (x_i^{(l)} - x_j^{(l)}) \cdot \varphi_{\hat{\mu}}(\mu_{i,j}).
 \end{align}
 
After extracting inter-message and intra-message, the aggregation function aggregates the neighbor messages of the node $i$. $\forall i \in \mathcal{V}_\mathcal{P} \cup \mathcal{V}_\mathcal{L}$:
 \begin{align}
    & m_i = \sum_{j \in \mathcal{N}(i)}m_{i,j}, \;
    \hat{m}_i = \sum_{j \in \mathcal{N}(i) }\frac{1}{d_{i,j}^{(l)} + 1} \cdot \hat{m}_{i,j}, \\
    & \mu_{i} = \sum_{j \in \mathcal{N}_*^{(l)}(i)} \varphi(\mu_{i,j}) \cdot \mu_{i,j}, \;
    \hat{\mu}_i = \sum_{j \in \mathcal{N}_*^{(l)}(i) }\frac{1}{d_{i,j}^{(l)} + 1} \cdot \hat{\mu}_{i,j},
 \end{align}
where $\mathcal{N}(i)$ is the neighbor of node $i$ in the same graph, and $\mathcal{N}_*^{(l)}(i) $ is the set of nodes associated with node $i$ in the other graph.

Finally, the update function updates the position and features of each node: 
 \begin{align}
    & x_i^{(l+1)}
    = \eta x_i^{(0)} + (1 - \eta) x_i^{(l)} + \hat{m}_i + \hat{\mu}_i, \; \forall i \in \mathcal{V}_\mathcal{P} \cup \mathcal{V}_\mathcal{L}, \\
    & h_i^{(l+1)} = (1 - \beta) \cdot h_i^{(l)} + \beta \cdot \varphi(h_i^{(l)}, m_i, \mu_i, h_i^{(0)}), \; \forall i \in \mathcal{V}_\mathcal{P} \cup \mathcal{V}_\mathcal{L},
\end{align}
where $\beta$ and $\eta$ are feature skip connection weight and coordinates skip connection weight, respectively. Through such a message-passing paradigm, our \refinemodel layers make to update coordinates iteratively.

\subsubsection{Fast Structure Correction}
Lastly, as \refinemodel updates structures by modifying the coordinates rather than the torsional angles, as is done in methods like DiffDock~\cite{Corso2022DiffDockDS} and other sampling-based methods, it is crucial to ensure the plausibility of bond lengths and bond angles of the updated structure $\hat{\mathcal{X}}^\mathcal{L}$. Therefore, fast structure correction steps, torsion alignment, and SMINA-based energy minimization are designed.

\textbf{Torsion Alignment.}
We employ a rapid torsion alignment for the updated structure.
The target of this alignment is to align the input structure $\mathcal{X}^\mathcal{L}$ with the updated structures $\hat{\mathcal{X}}^\mathcal{L}$ by rotating its torsional bonds.
Formally, let $(b_i, c_i)$ denote a $i$-th rotatable bond, where $b_i$ and $c_i$ are the starting and ending atoms of the bond, respectively. 
We randomly select a neighboring atom $a_i$ of $b_i$ and a neighbor atom $d_i$ of $c_i$ to calculate the dihedral angle $\hat{\delta}_i = \angle(a_ib_ic_i,b_ic_id_i)$ based on updated structure coordinates $\hat{\mathcal{X}}^\mathcal{L}$. Subsequently, we rotate the rotatable bond $(b_i, c_i)$ of input structures to match its dihedral angle $\delta_i$ the same as $\hat{\delta}_i$. This simple operation can be implemented efficiently using RDKit. 
After all rotatable bonds have been rotated, we align the rotated input structure to the updated structures to obtain the torsionally aligned structure $\hat{\mathcal{X}}^{\mathcal{L}\mathcal{T}}$. This process ensures the plausibility of bond lengths and bond angles in the torsionally aligned structure $\hat{\mathcal{X}}^{\mathcal{L}\mathcal{T}}$.

\textbf{Energy Minimization.}
To further enhance the reliability of \modelname, we implement an energy minimization on the torsionally aligned structure $\hat{\mathcal{X}}^{\mathcal{L}\mathcal{T}}$, when an inter-molecular steric clash between the protein and ligand is detected. This energy minimization is conducted using SMINA~\cite{Koes2013LessonsLI}, as it is a highly efficient tool for this process
compared with specialized energy minimization tool OpenMM~\cite{Eastman2016OpenMM7R} (details see Appendix.\ref{apdsec:eff_comp_smina}).
The output structure of this process is $\hat{\mathcal{X}}^{\mathcal{L}{'}}$.

\subsection{Training and Inference}
\subsubsection{\sitemodel}
The training object $L$ is a contrastive object defined before (Eq.~\ref{eq:pocketalign_loss}). For a protein and its candidate pockets set $S = \{\varsigma_1, \varsigma_2, ...\}$, the positive pair is the target pocket-ligand pair and the negative pairs are other pocket-ligand pairs. The pocket-ligand pairs across different proteins are not used. When training, we calculate the minimum center distance ($DCC_{min}$) between all candidate pockets and the ligand. If $DCC_{min} \leq 5.0$ \AA , we add the ligand center into $S$ to assert the existence of positive pairs for every protein (details see Appendix.\ref{apdsec:cpla_exp_settings}). 

\subsubsection{\refinemodel}
We design a physics-informed loss function for the \refinemodel module for training. 
The coordinates $X^{a, \mathcal{L}}$ and $X^{r, \mathcal{L}}$ output by the last layer of atomic level and residue level are both employed in the computation of this loss. Formally, the loss function can be expressed as follows: \begin{equation}
    L =
     L_{inter} +
     \lambda_1 L_{intra} +
     \lambda_2 L_{vdw} +
     \lambda_3 L_{bound},
\end{equation}
where $\lambda$ are weight hyper-parameters.
Among the four components, inter-distance map loss $L_{inter}$ is responsible for the RMSD accuracy. Other three items, namely intra-distance map loss $L_{intra}$, vdw constraint loss $L_{vdw}$, and bound matrix constraint loss $L_{bound}$ are employed for physical validity.
When training and inferencing, we follow previous work~\cite{Jumper2021HighlyAP} and employ the recycling strategy (details see Appendix.\ref{apdsec:biegmn_exp_settings}).
\vspace{-0.1in}
\section{Experiments}
\label{sec:exp}

\subsection{Settings}
\textbf{Dataset.} We conduct experimetns on PDBbind~\cite{Liu2017ForgingTB} v2020 and PoseBusters~\cite{Buttenschoen2023PoseBustersAD} datasets in this work. 
Our model is trained on the PDBbind dataset, where the training, validation, and testing set are constructed based on the time split strategy used in previous work~\cite{Pei2023FABindFA}.
PoseBusters, which contains 428 carefully selected data released from 1 January 2021 to 30 May 2023, is directly adopted to evaluate the ability to predict physically valid poses.

\textbf{Evaluation.}
Root-mean-square-deviation (RMSD) and centroid distance (CD) are used to evaluate the docking accuracy of different docking methods, and the PoseBusters~\cite{Buttenschoen2023PoseBustersAD} test suite is employed to evaluate the performance of predicting physically valid poses.
Additionally, as pocket prediction plays an important role in our framework, the distance between the center of the predicted pocket and the center of the ground-truth ligand structure (DCC), and the volume coverage rate (VCR) are employed to evaluate the pocket prediction accuracy (details in Appendix.\ref{apd:exp_settings}). 

 \begin{table}[t]
 \centering
 \caption{Blind docking performance on the PDBbind dataset. All methods take RDKit-generated ligand structures and holo protein structures as input, trying to predict bound complex structures.  \modelname-SC refers to the model variant that generates structures without implementing fast structure correction. \modelname-Random refers to the model variant that generates structures without high-quality initial poses. The best results are {\bf bold}, and the second best results are \underline{underlined}.}
 \label{tbl:blind_docking_test_set}
\begin{adjustbox}{width=1.0\textwidth}
\begin{threeparttable}
\begin{tabular}{lc|cc|cc|cc|cc}
\toprule
\multirow{3}{*}{Method} &
\multirow{2}{*}{Time average} & 
\multicolumn{4}{c|}{Time Split (363)} & 
\multicolumn{4}{c}{Timesplit Unseen (142)}\\
& 
& \multicolumn{2}{c}{RMSD \% below} 
& \multicolumn{2}{c|}{Centroid \% below}  
& \multicolumn{2}{c}{RMSD \% below} 
& \multicolumn{2}{c}{Centroid \% below} \\
& Seconds & 2.0\AA  & 5.0\AA  & 2.0\AA  & 5.0\AA  & 2.0\AA  & 5.0\AA & 2.0\AA & 5.0\AA \\
\midrule
QVINA-W
& 49* & 20.9 & 40.2 & 41.0 & 54.6 & 15.3 & 31.9 & 35.4 & 47.9 \\
GNINA
& 393 & 21.2 & 37.1 & 36.0 & 52.0 & 13.9 & 27.8 & 25.7 & 39.5  \\
VINA
& 119* & 10.3 & 36.2 & 32.3 & 55.2 & 7.8 & 25.5 & 24.1 & 41.8  \\ 
SMINA
& 146* & 13.5 & 33.9 & 38.0 & 55.9 & 9.0 & 25.7 & 29.9 & 41.7 \\
GLIDE
& 1405* & 21.8 & 33.6 & 36.1 & 48.7 & 19.6 & 28.7 & 29.4 & 40.6 \\
DSDP 
& 1.22 & 40.2 & 59.0 & 59.5 & 78.2 & 37.3 & 54.9 & 55.6 & 71.8 \\ 
\midrule
EquiBind
& \textbf{0.03} & 5.5 & 39.1 & 40.0 & 67.5 & 0.7 & 18.8 & 16.7 & 43.8 \\
TANKBind
& 0.87 & 17.6 & 57.8 & 55.0 & 77.8 & 3.5 & 43.7 & 40.9 & 70.8 \\
DiffDock
& 80 & 36.0 & 61.7 & 62.9 & 80.2 & 17.2 & 42.3 & 43.3 & 62.6 \\
FABind
& \underline{0.12} & 33.1 & 64.2 & 60.8 & 80.2 & 19.4 & 60.4 & 57.6 & 75.7 \\
\midrule
\textbf{\modelname-SC}
& 2.58 & \textbf{47.9} & \textbf{68.0} & \textbf{70.0} & \textbf{83.2} & \textbf{40.8} & \underline{60.6} & \textbf{65.5} & \textbf{78.9} \\
\textbf{\modelname}
& 2.97 & \underline{47.4} & \underline{66.9} & \underline{66.7} & \textbf{83.2} & \textbf{40.8} & \textbf{61.3} & \underline{60.6} & \textbf{78.9} \\
\bottomrule
\end{tabular}
\begin{tablenotes}
 \item[1] The time of consumption is denoted with * if it only consumes CPU.
 \item[2] All results of baselines are taken from \cite{Pei2023FABindFA} for fair comparison.
 \end{tablenotes}
\end{threeparttable}
\end{adjustbox}

\end{table}

\subsection{Overall Performance on the PDBbind}
We first assess the comprehensive performance of \modelname on the PDBbind dataset, encompassing both blind docking and site-specific docking settings.

\subsubsection{Blind Docking}
As demonstrated in Table.\ref{tbl:blind_docking_test_set}, \modelname outperforms all baseline methods. Specifically, \modelname achieves a remarkable success rate of 47.4\% (where RMSD < 2.0 \AA), surpassing the previous SOTA GDL method, DiffDock, which has a success rate of 36.0\%.
Recent GPU-accelerated docking methods have also made significant progress in blind docking. However, when compared to DSDP, which is the top-performing sampling-based method in the PDBbind test set, \modelname still exhibits superior performance across all metrics. Notably, as elucidated in Section~\ref{sec:iterative_refine}, \modelname employs the same sampling algorithm as DSDP for generating the initial structure. Yet, our framework allows \modelname to significantly outperform DSDP.

Beyond accuracy, efficiency is a critical performance measure for molecular docking methods. As indicated in Table~\ref{tbl:blind_docking_test_set}, \modelname maintains a competitive level of efficiency, despite the inclusion of an energy minimization operation to enhance accuracy and reliability. Molecular docking methods invariably face a trade-off between efficiency and accuracy. However, the data presented in Table~\ref{tbl:blind_docking_test_set} suggest that \modelname could serve as a viable tool for practical applications, balancing these two crucial aspects effectively.

\subsubsection{Site-specific Docking}
\begin{figure}[t]
  \includegraphics[width=\linewidth]{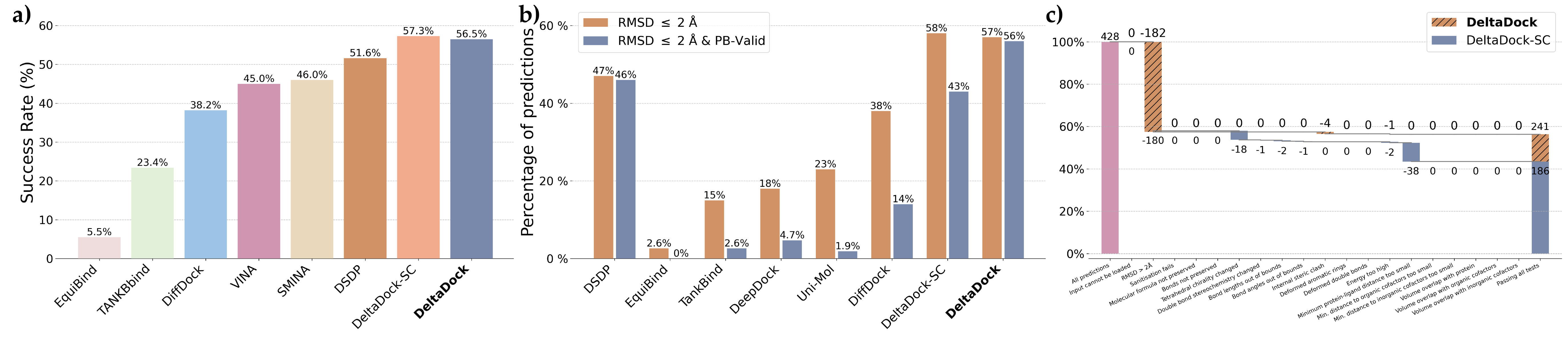}
  \caption{Site-specific docking performance.
  (a) Overall Performance of different methods on the PDBbind test set. The search space was delineated by extending the minimum and maximum of the x, y, and z coordinates of the ligand by 4 \AA \; respectively. 
  For TANKBind, we directly supply the protein block with a radius of 20~\AA \; centered around the ground-truth ligand center to the model. 
  (b) Overall performance of different methods on the PoseBusters dataset.
  (c) A waterfall plot for illustrating the PoseBusters tests as filters for both \modelname and \modelname-SC predictions. The evaluation results for \modelname are denoted above the lines, while those for \modelname-SC are annotated below.}
  \label{fig:site_specific_pdb_and_posebuster}
 \end{figure}

Most existing GDL methods, such as DiffDock and EquiBind, are primarily designed for blind docking scenarios and are not inherently suited for site-specific docking tasks. However, \modelname seamlessly integrates blind docking and site-specific docking settings. In this context, the pocket is directly provided, eliminating the need for pocket selection via \sitemodel.
The performance of \modelname in site-specific docking is illustrated in Fig.\ref{fig:site_specific_pdb_and_posebuster}. When supplied with predefined binding sites, traditional sampling methods exhibit a significant improvement in results. For instance, the docking success rate of VINA escalates from 10.3\% to 45.0\%. Despite this enhancement, \modelname consistently surpasses all baselines.
Previous research suggested that while GDL docking methods excel at pocket searching, traditional methods tend to outperform GDL models in site-specific docking tasks~\cite{Yu2023DoDL}. However, as evidenced by the results presented in Table.\ref{tbl:blind_docking_test_set} and Fig.\ref{fig:site_specific_pdb_and_posebuster}, \modelname exhibits superior performance in both blind and site-specific docking scenarios, demonstrating its versatility and robustness in handling diverse docking settings.

\subsection{Evaluation of Generalization Capability}
Historically, GDL docking methods have demonstrated limited generalization capabilities. Here, we first examine the blind docking performance of \modelname and baseline methods on the unseen set of the PDBbind test, following prior work.
As indicated in Table~\ref{tbl:blind_docking_test_set}, the docking success rate of all methods on the unseen set from the PDBbind test is generally lower than that on the complete PDBbind test set. For example, the performance of GLIDE and QVINA-W shows a modest decline of 2.2\% and 5.6\%, respectively. For GDL baselines, the performance decrement is more pronounced. Notably, TANKBind and the SOTA GDL method DiffDock experience a performance drop of 14.1\% and 18.8\%. This outcome suggests that the unseen test set is more challenging than the whole test set. However, \modelname demonstrates competitive performance, achieving a docking success rate of 40.8\%. Compared to FABind, the best-performing GDL baseline on the unseen test set, \modelname surpasses it by a significant 20.1\% in terms of docking success rate.

\begin{figure}[t]
  \includegraphics[width=1.0\linewidth]{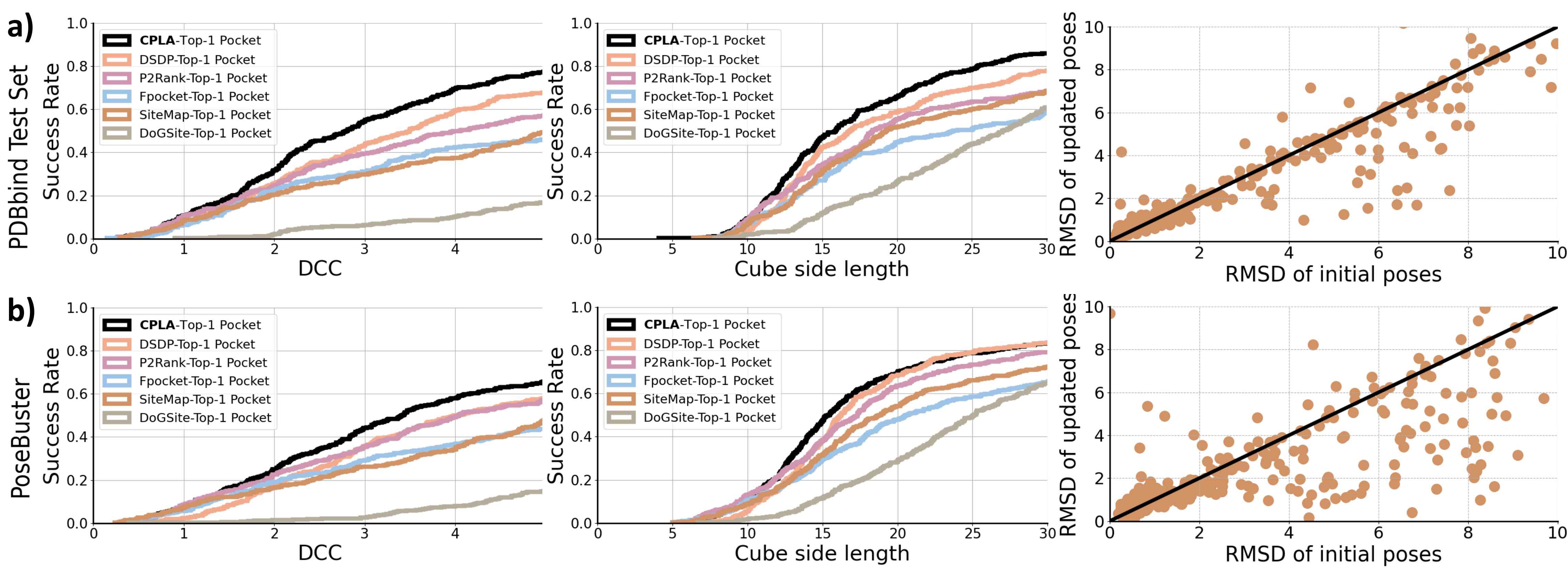}
  \caption{Further analysis on the (a) PDBbind and (b) PoseBusters dataset.
  Left: DCC cumulative curve of top-1 pockets.
  Middle: VCR cumulative curve of top-1 pockets.
  Right: Scatter plot of RMSD of initial and updated poses.
  All experiments are conducted in the blind docking setting.}
  \label{fig:further_analysis}
  \vspace{-0.2cm}
 \end{figure}

\vspace{-0.2cm}
\subsection{Evaluation of Pose Validity}
We further investigate \modelname's ability to predict physically valid structures by employing the PoseBusters test suite, as designed by Buttenschoen et al.~\cite{Buttenschoen2023PoseBustersAD}. In addition to the RMSD between predicted and ground-truth poses, the test suite incorporates 18 checks, encompassing chemical validity and consistency, intramolecular validity, and intermolecular validity.
When physical validity is considered, the docking success rates of traditional sampling methods remain stable, while the performance of previous geometric deep learning methods significantly declines, especially for TANKBind, DeepDock, and Uni-Mol.
The \modelname-SC variant, even without the application of the fast structure correction step, shows significant improvement over previous methods.
These results substantiate \modelname's capacity to predict physically valid structures, thereby affirming its reliability for practical applications.

\vspace{-0.2cm}
\subsection{Further Analysis}
\vspace{-0.1cm}
\subsubsection{Pocket-ligand Alignment and Iterative Refinement}
Beyond the overall docking performance, the pocket-ligand alignment and iterative refinement results are explored~(Fig.~\ref{fig:further_analysis}). As depicted in the figure, \sitemodel predicts significantly more accurate pockets than other methods and \refinemodel can diminish the discrepancy between ground-truth structures and input structures. Generally, the PDBbind test set poses a more significant challenge to \refinemodel than the PoseBusters dataset. And for \sitemodel, PoseBusters dataset is more challenging otherwise.
The consistent good performance on the two datasets demonstrates the effectiveness and generalization capacity of \sitemodel and \refinemodel.

 \begin{wraptable}{r}{0.5\textwidth}
 \centering\vspace{-3em}
 \caption{Results of ablation study.}
 \label{tbl:ablation_study}
\begin{threeparttable}
\begin{tabular}{l|cc}
\toprule
\multirow{2}{*}{Method} &
\multicolumn{2}{c}{RMSD \% below 2 \AA } \\
& PDBbind
& PoseBusters \\ 
\midrule
\textbf{\modelname} & \textbf{47.4} & \textbf{49.3} \\
\midrule
w/o \sitemodel & 41.2 & 43.7 \\
w/o \refinemodel & 44.6 & 41.8 \\
\midrule
w/o Residue Level  & 44.6 & 44.4 \\
w/o Atom Level & 44.6 & 42.1 \\
\bottomrule
\end{tabular}
\end{threeparttable}
\end{wraptable}

\vspace{-0.1cm}
\subsubsection{Ablation Studies}
In this section, ablation studies are conducted to assess the contributions of different components. 
We first ablate the whole \sitemodel or \refinemodel, and then the residue-level or the atom-level in \refinemodel (see Appendix.~\ref{apdsec:ablation_settings} for implement details).
As illustrated in Table~\ref{tbl:ablation_study}, it becomes clear that each component, encompassing \sitemodel and the bi-level strategy in \refinemodel, plays a significant role in enhancing the overall performance of \modelname. 
Due to the space limitation, a full ablation study can be found in Appendix.~\ref{apdsec:full_ablation}.

\vspace{-0.1in}
\section{Conclusion}
\label{sec:conclusion}
In this work, we proposed \modelname, a unified framework for accurate, efficient, and physically reliable molecular docking.
\modelname was a two-stage docking framework, consisting of pocket prediction and site-specific docking.
We innovatively reframed the pocket prediction task as a pocket-ligand alignment problem and then followed a hybrid strategy to jointly utilize both GDL and physics-informed traditional algorithms for site-specific docking. 
Comprehensive experiments demonstrated the superior performance of \modelname.
Notably, in the blind docking setting, \modelname achieved a 31\% relative improvement over the docking success rate compared with the previous state-of-the-art GDL model.
We hope this work will further facilitate the broad application and continued development of the molecular docking framework.

\section{Acknowledgements}
We extend our gratitude to the reviewers for their valuable and insightful feedback, which significantly improved this work. We are also grateful to Lixue Cheng from Microsoft Research Asia for her helpful suggestions and comments.
This research was supported by grants from the National Natural Science Foundation of China (Grant No. 623B2095) and the Fundamental Research Funds for the Central Universities.

\newpage
\bibliographystyle{unsrt}
\bibliography{sample-base}

\begin{thebibliography}{10}

\bibitem{Bronstein2021GeometricDL}
Michael~M. Bronstein, Joan Bruna, Taco Cohen, and Petar Veličković.
\newblock Geometric deep learning: Grids, groups, graphs, geodesics, and gauges.
\newblock In {\em ArXiv}, 2021.

\bibitem{zhang2023fullatom}
Zaixi Zhang, Zepu Lu, Zhongkai Hao, Marinka Zitnik, and Qi~Liu.
\newblock Full-atom protein pocket design via iterative refinement.
\newblock In {\em NeurIPS’23}, 2023.

\bibitem{Zhang2023LearningSP}
Zaixi Zhang and Qi~Liu.
\newblock Learning subpocket prototypes for generalizable structure-based drug design.
\newblock In {\em ICML'24}, 2023.

\bibitem{Du2016InsightsIP}
Xing Du, Yi~Li, Yuan-Ling Xia, Shi-Meng Ai, Jing Liang, Peng Sang, Xing lai Ji, and Shu-Qun Liu.
\newblock Insights into protein–ligand interactions: Mechanisms, models, and methods.
\newblock {\em International Journal of Molecular Sciences}, 17:144, 2016.

\bibitem{Li2021StructureawareIG}
Shuangli Li, Jingbo Zhou, Tong Xu, Liang Huang, Fan Wang, Haoyi Xiong, Weili Huang, Dejing Dou, and Hui Xiong.
\newblock Structure-aware interactive graph neural networks for the prediction of protein-ligand binding affinity.
\newblock In {\em KDD'21}, 2021.

\bibitem{Lyu2019UltralargeLD}
Jiankun Lyu, Sheng Wang, Trent~E. Balius, Isha Singh, Anat Levit, Yurii~S. Moroz, Matthew~J. O’Meara, Tao Che, Enkhjargal Algaa, Kateryna~A Tolmachova, Andrey~A. Tolmachev, Brian~K. Shoichet, Bryan~L. Roth, and John~J. Irwin.
\newblock Ultra-large library docking for discovering new chemotypes.
\newblock {\em Nature}, 566:224 -- 229, 2019.

\bibitem{Bender2021APG}
Brian~Joseph Bender, Stefan Gahbauer, Andreas Luttens, Jiankun Lyu, Chase~M Webb, Reed~M. Stein, Elissa~A. Fink, Trent~E. Balius, Jens Carlsson, John~J. Irwin, and Brian~K. Shoichet.
\newblock A practical guide to large-scale docking.
\newblock {\em Nature protocols}, page 4799–4832, 2021.

\bibitem{Strk2022EquiBindGD}
Hannes St{\"a}rk, Octavian-Eugen Ganea, Lagnajit Pattanaik, Regina Barzilay, and T.~Jaakkola.
\newblock Equibind: Geometric deep learning for drug binding structure prediction.
\newblock In {\em ICML'22}, 2022.

\bibitem{Corso2022DiffDockDS}
Gabriele Corso, Hannes St{\"a}rk, Bowen Jing, Regina Barzilay, and Tommi~S. Jaakkola.
\newblock Diffdock: Diffusion steps, twists, and turns for molecular docking.
\newblock In {\em ICLR'23}, 2023.

\bibitem{Zhang2023ASS}
Zaixi Zhang, Jiaxian Yan, Qi~Liu, and Enhong Chen.
\newblock A systematic survey in geometric deep learning for structure-based drug design.
\newblock {\em ArXiv}, abs/2306.11768, 2023.

\bibitem{Pei2023FABindFA}
Qizhi Pei, Kaiyuan Gao, Lijun Wu, Jinhua Zhu, Yingce Xia, Shufang Xie, Tao Qin, Kun He, Tie-Yan Liu, and Rui Yan.
\newblock Fabind: Fast and accurate protein-ligand binding.
\newblock In {\em NeurIPS'23}, 2023.

\bibitem{Zhang2022E3BindAE}
Yangtian Zhang, Huiyu Cai, Chence Shi, and Jian Tang.
\newblock E3bind: An end-to-end equivariant network for protein-ligand docking.
\newblock In {\em ICLR'23}, 2023.

\bibitem{Hassan2017ProteinLigandBD}
Nafisa Hassan, Amr Alhossary, Yuguang Mu, and C.~Kwoh.
\newblock Protein-ligand blind docking using quickvina-w with inter-process spatio-temporal integration.
\newblock {\em Scientific Reports}, 7, 2017.

\bibitem{Zhou2023UniMolAU}
Gengmo Zhou, Zhifeng Gao, Qiankun Ding, Hang Zheng, Hongteng Xu, Zhewei Wei, Linfeng Zhang, and Guolin Ke.
\newblock Uni-mol: A universal 3d molecular representation learning framework.
\newblock In {\em ICLR'23}, 2023.

\bibitem{Zhu2023DiffBindFRAS}
Jintao Zhu, Zhonghui Gu, Jianfeng Pei, and Luhua Lai.
\newblock Diffbindfr: An se(3) equivariant network for flexible protein-ligand docking.
\newblock In {\em ArXiv}, 2023.

\bibitem{Buttenschoen2023PoseBustersAD}
Martin Buttenschoen, Garrett~M. Morris, and Charlotte~M. Deane.
\newblock Posebusters: Ai-based docking methods fail to generate physically valid poses or generalise to novel sequences.
\newblock {\em Chemical Science}, 2023.

\bibitem{Reidenbach2023CoarsenConfEC}
Danny Reidenbach and Aditi~S. Krishnapriyan.
\newblock Coarsenconf: Equivariant coarsening with aggregated attention for molecular conformer generation.
\newblock In {\em ArXiv}, 2023.

\bibitem{Trott2010AutoDockVI}
Oleg Trott and Arthur~J. Olson.
\newblock Autodock vina: Improving the speed and accuracy of docking with a new scoring function, efficient optimization, and multithreading.
\newblock {\em Journal of Computational Chemistry}, 31:455--461, 2010.

\bibitem{Koes2013LessonsLI}
David~Ryan Koes, Matthew~P. Baumgartner, and Carlos~J. Camacho.
\newblock Lessons learned in empirical scoring with smina from the csar 2011 benchmarking exercise.
\newblock {\em Journal of chemical information and modeling}, 53 8:1893--904, 2013.

\bibitem{Nocedal2000NumericalO}
Jorge Nocedal and Stephen~J. Wright.
\newblock {\em Numerical Optimization}.
\newblock Springer New York, NY, 2000.

\bibitem{Ding2023VinaGPU2F}
Ji~Ding, Shi xiong Tang, Zheming Mei, Lingyue Wang, Qinqin Huang, Haifeng Hu, Ming Ling, and Jiansheng Wu.
\newblock Vina-gpu 2.0: Further accelerating autodock vina and its derivatives with graphics processing units.
\newblock {\em Journal of chemical information and modeling}, 63:1982–1998, 2023.

\bibitem{Yu2023UniDockGD}
Yuejiang Yu, Chun Cai, Jiayue Wang, Zonghua Bo, Zhengdan Zhu, and Hang Zheng.
\newblock Uni-dock: Gpu-accelerated docking enables ultralarge virtual screening.
\newblock {\em Journal of chemical theory and computation}, 19:3336--3345, 2023.

\bibitem{Huang2023DSDPAB}
Yupeng Huang, Hong Zhang, Siyuan Jiang, Dajiong Yue, Xiaohan Lin, Jun Zhang, and Yi~Qin Gao.
\newblock Dsdp: A blind docking strategy accelerated by gpus.
\newblock {\em Journal of chemical information and modeling}, 63:4355–4363, 2023.

\bibitem{Guilloux2009FpocketAO}
Vincent~Le Guilloux, Peter Schmidtke, and Pierre Tuff{\'e}ry.
\newblock Fpocket: An open source platform for ligand pocket detection.
\newblock {\em BMC Bioinformatics}, 10:168 -- 168, 2009.

\bibitem{Krivk2018P2RankML}
Radoslav Kriv{\'a}k and David Hoksza.
\newblock P2rank: machine learning based tool for rapid and accurate prediction of ligand binding sites from protein structure.
\newblock {\em Journal of Cheminformatics}, 10:39, 2018.

\bibitem{Kandel2021PUResNetPO}
Jeevan Kandel, Hilal Tayara, and Kil to~Chong.
\newblock Puresnet: prediction of protein-ligand binding sites using deep residual neural network.
\newblock {\em Journal of Cheminformatics}, 13:65, 2021.

\bibitem{greg_landrum_2022_6798971}
Greg Landrum, Paolo Tosco, Brian Kelley, Ric, sriniker, gedeck, Riccardo Vianello, NadineSchneider, Eisuke Kawashima, Andrew Dalke, David Cosgrove, Dan N, Gareth Jones, Brian Cole, Matt Swain, Samo Turk, AlexanderSavelyev, Alain Vaucher, Maciej Wójcikowski, Ichiru Take, Daniel Probst, Kazuya Ujihara, Vincent~F. Scalfani, guillaume godin, Axel Pahl, Francois Berenger, JLVarjo, strets123, JP, and DoliathGavid.
\newblock rdkit/rdkit: 2022\_03\_4 (q1 2022) release, July 2022.

\bibitem{Xiong2020PushingTB}
Zhaoping Xiong, Dingyan Wang, Xiaohong Liu, Feisheng Zhong, Xiaozhe Wan, Xutong Li, Zhaojun Li, Xiaomin Luo, Kaixian Chen, Hualiang Jiang, and Mingyue Zheng.
\newblock Pushing the boundaries of molecular representation for drug discovery with graph attention mechanism.
\newblock {\em Journal of medicinal chemistry}, 63:8749–8760, 2020.

\bibitem{Jing2020LearningFP}
Bowen Jing, Stephan Eismann, Patricia Suriana, Raphael J.~L. Townshend, and Ron~O. Dror.
\newblock Learning from protein structure with geometric vector perceptrons.
\newblock In {\em ICLR '21}, 2020.

\bibitem{Lu2022TANKBindTN}
Wei Lu, Qifeng Wu, Jixian Zhang, Jiahua Rao, Chengtao Li, and Shuangjia Zheng.
\newblock Tankbind: Trigonometry-aware neural networks for drug-protein binding structure prediction.
\newblock In {\em NeurIPS'22}, 2022.

\bibitem{Gilmer2017NeuralMP}
Justin Gilmer, Samuel~S. Schoenholz, Patrick~F. Riley, Oriol Vinyals, and George~E. Dahl.
\newblock Neural message passing for quantum chemistry.
\newblock In {\em ICML'17}, 2017.

\bibitem{Eastman2016OpenMM7R}
Peter~K. Eastman, Jason~M. Swails, John~D. Chodera, Robert~T. McGibbon, Yutong Zhao, Kyle~A. Beauchamp, Lee‐Ping Wang, Andrew~C. Simmonett, Matthew~P. Harrigan, Chaya~D. Stern, Rafal~P. Wiewiora, Bernard~R. Brooks, and Vijay~S. Pande.
\newblock Openmm 7: Rapid development of high performance algorithms for molecular dynamics.
\newblock {\em PLoS Computational Biology}, 13, 2016.

\bibitem{Jumper2021HighlyAP}
John~M. Jumper, Richard Evans, Alexander Pritzel, Tim Green, Michael Figurnov, Olaf Ronneberger, Kathryn Tunyasuvunakool, Russ Bates, Augustin Z{\'i}dek, Anna Potapenko, Alex Bridgland, Clemens Meyer, Simon A~A Kohl, Andy Ballard, Andrew Cowie, Bernardino Romera-Paredes, Stanislav Nikolov, Rishub Jain, Jonas Adler, Trevor Back, Stig Petersen, David~A. Reiman, Ellen Clancy, Michal Zielinski, Martin Steinegger, Michalina Pacholska, Tamas Berghammer, Sebastian Bodenstein, David Silver, Oriol Vinyals, Andrew~W. Senior, Koray Kavukcuoglu, Pushmeet Kohli, and Demis Hassabis.
\newblock Highly accurate protein structure prediction with alphafold.
\newblock {\em Nature}, 596:583 -- 589, 2021.

\bibitem{Liu2017ForgingTB}
Zhihai Liu, Minyi Su, Li~Han, Jie Liu, Qifan Yang, Yan Li, and Renxiao Wang.
\newblock Forging the basis for developing protein-ligand interaction scoring functions.
\newblock {\em Accounts of chemical research}, 50 2:302--309, 2017.

\bibitem{Yu2023DoDL}
Yuejiang Yu, Shuqi Lu, Zhifeng Gao, Hang Zheng, and Guolin Ke.
\newblock Do deep learning models really outperform traditional approaches in molecular docking?
\newblock In {\em ArXiv}, 2023.

\bibitem{Wang2022LearningHP}
Limei Wang, Haoran Liu, Yi~Liu, Jerry Kurtin, and Shuiwang Ji.
\newblock Learning hierarchical protein representations via complete 3d graph networks.
\newblock In {\em ICLR'23}, 2022.

\bibitem{Fu2022SIPFSM}
Tianfan Fu and Jimeng Sun.
\newblock Sipf: Sampling method for inverse protein folding.
\newblock In {\em KDD'22}, 2022.

\bibitem{Dohan2021ImprovingPF}
David Dohan, Andreea Gane, Maxwell~L. Bileschi, David Belanger, and Lucy~J. Colwell.
\newblock Improving protein function annotation via unsupervised pre-training: Robustness, efficiency, and insights.
\newblock In {\em KDD'21}, 2021.

\bibitem{graef2023binding}
Joel Graef, Christiane Ehrt, and Matthias Rarey.
\newblock Binding site detection remastered: Enabling fast, robust, and reliable binding site detection and descriptor calculation with dogsite3.
\newblock {\em Journal of Chemical Information and Modeling}, 63(10):3128--3137, 2023.

\bibitem{halgren2007new}
Tom Halgren.
\newblock New method for fast and accurate binding-site identification and analysis.
\newblock {\em Chemical biology \& drug design}, 69(2):146--148, 2007.

\bibitem{halgren2009identifying}
Thomas~A Halgren.
\newblock Identifying and characterizing binding sites and assessing druggability.
\newblock {\em Journal of chemical information and modeling}, 49(2):377--389, 2009.

\bibitem{OBoyle2011OpenBA}
Noel~M. O'Boyle, Michaela~S. Banck, Craig~A. James, Chris Morley, Tim Vandermeersch, and Geoffrey~R. Hutchison.
\newblock Open babel: An open chemical toolbox.
\newblock {\em Journal of Cheminformatics}, 3:33 -- 33, 2011.

\bibitem{lin2022language}
Zeming Lin, Halil Akin, Roshan Rao, Brian Hie, Zhongkai Zhu, Wenting Lu, Nikita Smetanin, Allan dos Santos~Costa, Maryam Fazel-Zarandi, Tom Sercu, Sal Candido, et~al.
\newblock Language models of protein sequences at the scale of evolution enable accurate structure prediction.
\newblock {\em bioRxiv}, 2022.

\bibitem{MndezLucio2021AGD}
Oscar M{\'e}ndez-Lucio, Mazen Ahmad, Ehecatl~Antonio del Rio-Chanona, and J{\"o}rg~Kurt Wegner.
\newblock A geometric deep learning approach to predict binding conformations of bioactive molecules.
\newblock {\em Nat. Mach. Intell.}, 3:1033--1039, 2021.

\bibitem{Meli2020spyrmsdSR}
Rocco Meli and Philip~Charles Biggin.
\newblock spyrmsd: symmetry-corrected rmsd calculations in python.
\newblock {\em Journal of Cheminformatics}, 12, 2020.

\bibitem{Kingma2015AdamAM}
Diederik~P. Kingma and Jimmy Ba.
\newblock Adam: A method for stochastic optimization.
\newblock In {\em ICLR'15}, 2015.

\bibitem{Gao2023DrugCLIPCP}
Bowen Gao, Bo~Qiang, Haichuan Tan, Minsi Ren, Yinjun Jia, Minsi Lu, Jingjing Liu, Weiying Ma, and Yanyan Lan.
\newblock Drugclip: Contrastive protein-molecule representation learning for virtual screening.
\newblock In {\em NeurIPS '23}, 2023.

\bibitem{landrum2013rdkit}
Greg Landrum et~al.
\newblock Rdkit: A software suite for cheminformatics, computational chemistry, and predictive modeling.
\newblock {\em Greg Landrum}, 2013.

\bibitem{Abramson2024AccurateSP}
Josh Abramson, Jonas Adler, Jack Dunger, Richard Evans, Tim Green, Alexander Pritzel, Olaf Ronneberger, Lindsay Willmore, Andrew~J Ballard, Joshua Bambrick, Sebastian~W Bodenstein, David~A Evans, Chia-Chun Hung, Michael O’Neill, David Reiman, Kathryn Tunyasuvunakool, Zachary Wu, Akvilė Žemgulytė, Eirini Arvaniti, Charles Beattie, Ottavia Bertolli, Alex Bridgland, Alexey Cherepanov, Miles Congreve, Alexander~Imani Cowen-Rivers, Andrew Cowie, Michael Figurnov, Fabian~B Fuchs, Hannah Gladman, Rishub Jain, Yousuf~A Khan, Caroline M~R Low, Kuba Perlin, Anna Potapenko, Pascal Savy, Sukhdeep Singh, Adrian Stecula, Ashok Thillaisundaram, Catherine Tong, Sergei Yakneen, Ellen~D. Zhong, Michal Zielinski, Augustin Ž{\'i}dek, Vic-613 tor Bapst, Pushmeet Kohli, Max Jaderberg, Demis Hassabis, and John~M. Jumper.
\newblock Accurate structure prediction of biomolecular interactions with alphafold3.
\newblock {\em Nature}, 2024.

\end{thebibliography}

\appendix
\clearpage
\makeatletter 
\makeatother
\renewcommand{\appendixname}{Appendix~\Alph{section}}
 
\section{More Detailed Descriptions}

\subsection{Dataset Preprocessing}
We follow the time split strategy used in previous work~\cite{Strk2022EquiBindGD, Lu2022TANKBindTN, Pei2023FABindFA} to split the dataset to construct the train, validation, and test set.
All compounds discovered in or after 2019 are in the test and validation sets, and only those found before 2019 are in the training set. The training set, validation set, and test set have 17,299, 968, and 363 complexes, respectively.
The overall performance of docking methods is evaluated on the time spit test set following previous works.
In this work, we only select the protein chains within 10 \AA \; to the ligand structure.

\subsection{Dataset Statistics}
\label{apdsec:data_statistics}
Proteins are inherently macromolecules composed of multiple chains, with each chain potentially containing hundreds or even thousands of residues~\cite{Wang2022LearningHP, Fu2022SIPFSM, Dohan2021ImprovingPF}.
In Table.\ref{apd:pdb_statistics}, we statistically analyze the PDBbind time-split test set and count atom numbers in proteins. Notably, it can be observed that the number of atoms escalates substantially as the cutoff value increases.

\begin{table}[ht]
\caption{Statistics of the PDBbind time split test set.}
\label{apd:pdb_statistics}
\begin{adjustbox}{width=1.0\textwidth}
\begin{tabular}{l|cc|cc}
\toprule

\multirow{2}{*}{\textbf{Data}} &
\multicolumn{2}{c}{\textbf{Average}} 
& \multicolumn{2}{c}{\textbf{Maximum}}  \\
& Number of $C_\alpha$ 
& Number of atoms
& Number of $C_\alpha$ 
& Number of atoms \\
\midrule
Entire protein structure
& 322 & 2,536 & 1,488 & 11,697 \\
Structure within 40.0 \AA \; cubic box centered on the ligand
& 179 & 1,602 & 400 & 3,055\\
Structure within 15.0 \AA \; from ligand
& 111 & 1,050 & 213 & 1,944\\
Structure within 12.0 \AA \; from ligand 
& 73 & 740 & 164 & 1,582 \\
Structure within 8.0 \AA \; from ligand
& 30 & 379 & 75 & 986 \\
Structure within 6.0 \AA \; from ligand 
& 16 & 207 & 45 & 548 \\
\bottomrule
\end{tabular}
\end{adjustbox}
\end{table}

\subsection{Example of Large Pocket}
\label{apdsec:large_pocket_exmaple}
\begin{figure}[ht]
  \centering
  \includegraphics[width=0.7\linewidth]{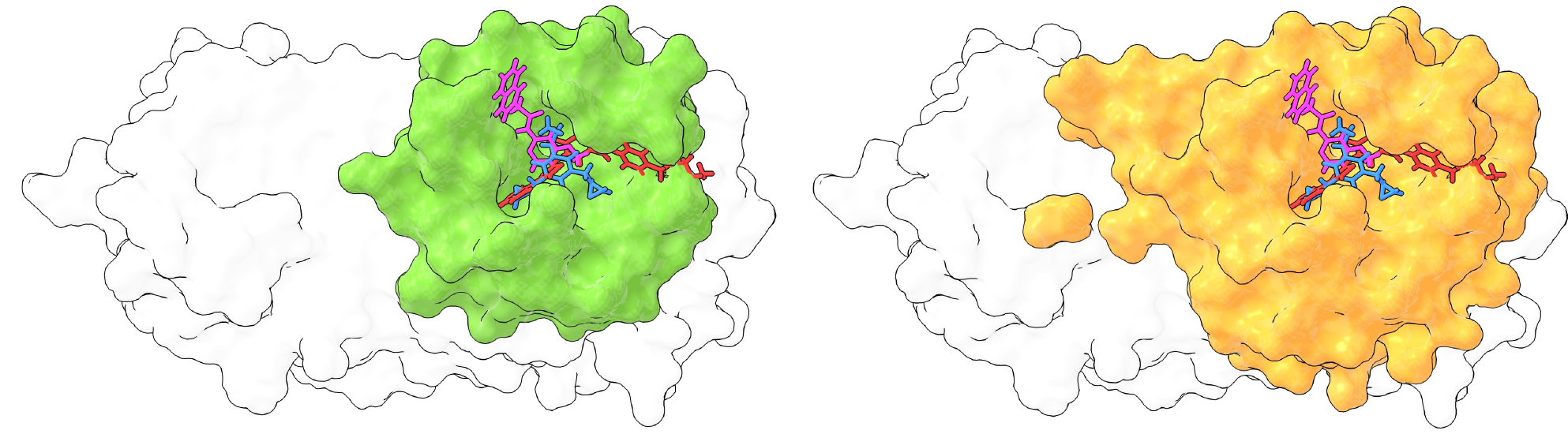}
  \caption{
  The main protease of SARS-CoV-2 is depicted by the white surface. The ligand structures in pink, blue, and red correspond to PDB 5RGY, 7AQJ, and 7JU7, respectively. 
  \textbf{Left:} The green pocket, a protein structure truncated to within 12.0 \AA \; of the blue structure, is insufficient to encompass the pocket structure necessary for predicting the red structure. 
  \textbf{Right:} The orange pocket, truncated within a 40.0 \AA \; box utilized by \modelname, is ample to cover the entire pocket.}
  \label{apdfig:bilevel_example}
 \end{figure}

Large pockets that consist of several sub-pockets generally exist. For example, the main protease of SARS-CoV-2 (Fig.~\ref{apdfig:bilevel_example}).
 
\subsection{Analysis of Existing Pocket Prediction Methods}
\label{apdsec:analys_pocket_pred}
\begin{figure}[ht]
 \centering
  \includegraphics[width=0.7\linewidth]{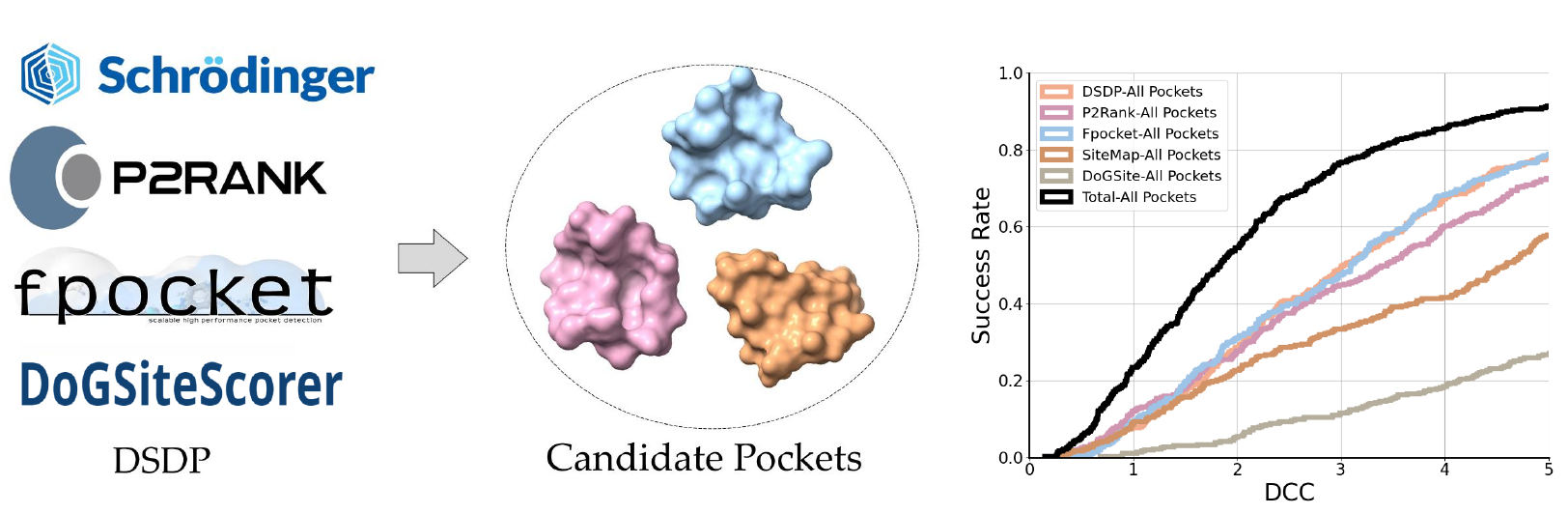}
  \caption{
  Performance of different pocket prediction methods on the PDBbind test set. The hit rate is significantly improved by ensembling the predicted pockets from various methods.}
  \label{fig:apd_dcc_total}
 \end{figure}

As depicted in Fig.\ref{fig:apd_dcc_total}, existing pocket prediction methods generally achieve a hit rate of approximately 70\%-80\%, where the distance between the predicted pocket center and ligand center (DCC) is less than 5.0 \AA. Notably, when leveraging combined predictions from multiple methods, the hit rate significantly increases to nearly 95\%.
Motivated by this observation, \modelname begins with a ready-to-dock ligand and a candidate pocket set derived from a suite of existing pocket prediction models. 

\begin{figure}[h]
 \centering
  \includegraphics[width=0.7\linewidth]{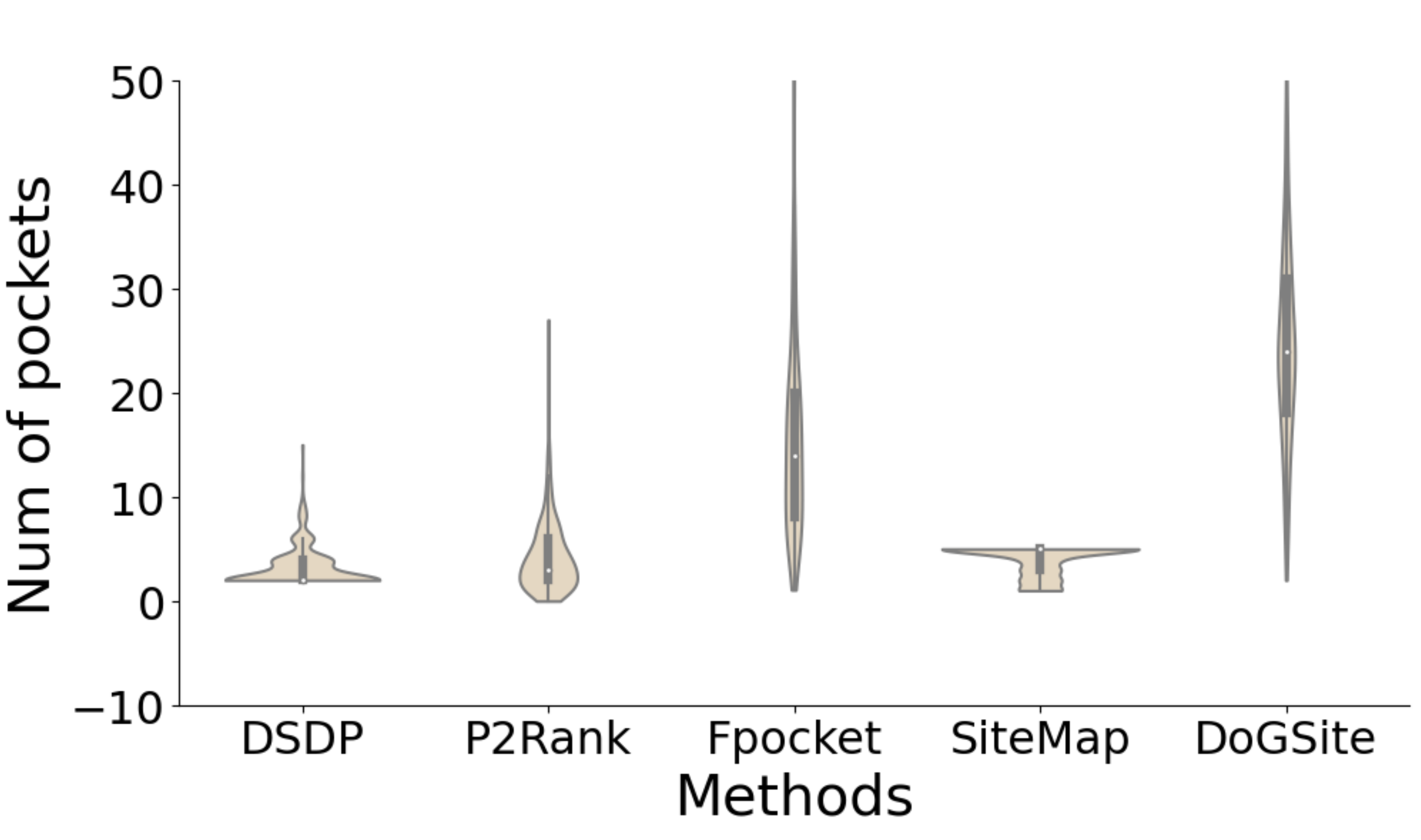}
  \caption{
  Pocket numbers violin plot of different methods.
  Pocket prediction methods generally predict a series of druggable pockets.}
  \label{fig:apd_pocket_num}
 \end{figure}

We further statistics how many pockets these methods predict in Fig.~\ref{fig:apd_pocket_num}. We observe that Fpocket~\cite{Guilloux2009FpocketAO}, and DoGSite3~\cite{graef2023binding} output much more pockets than DSDP~\cite{Huang2023DSDPAB}, P2Rank~\cite{Krivk2018P2RankML}, and SiteMap~\cite{halgren2007new, halgren2009identifying}. Combining information from Fig.~\ref{fig:apd_pocket_num} and Fig.\ref{fig:apd_dcc_total}, it is evident that the pockets predicted by DSDP and P2rank are highly druggable. Other methods, in contrast,  tend to predict many non-druggable pockets.

\subsection{Efficiency Comparison between SMINA and OpenMM}
\label{apdsec:eff_comp_smina}
For AI-based structure prediction methods, including AlphaFold2~\cite{Jumper2021HighlyAP}, it is common practice to employ energy minimization methods for post-processing to ensure the physical validity of the predicted structures.
While specialized methods like OpenMM are available for energy minimization, we opted not to use them due to computational efficiency considerations. Specifically, we found that SMINA, which is typically known as a docking method, requires only approximately 0.4 seconds for energy minimization. This is significantly faster than methods like OpenMM, which can take several minutes to tens of minutes per protein-ligand pair, as illustrated in the Table.~\ref{apdtbl:pdb_eff_compare} below.

For molecular docking, efficiency is crucial, and specialized methods such as OpenMM can be excessively time-consuming. What's more, it is important to note that SMINA, although generally regarded as a docking method, is not employed for docking in our workflow but rather utilized in its minimization mode for energy minimization. 

\begin{table}[ht]
\centering
\caption{Efficiency Comparison between SMINA and OpenMM.}
\label{apdtbl:pdb_eff_compare}
\begin{adjustbox}{width=0.5\textwidth}
\begin{tabular}{l|c}
\toprule
\textbf{Methods} &
\textbf{Time (per protein-ligand pair)}\\
\midrule
SMINA & about 0.4 seconds \\
OpenMM & several minutes to tens of minutes \\
\bottomrule
\end{tabular}
\end{adjustbox}
\end{table}

\subsection{Graph Construction}
\label{apd:graph_construct}
\textbf{Ligand Graph.}
The input ligand $L$ is first represented as a ligand graph $\mathcal{G}^\mathcal{L}=(\mathcal{V}^\mathcal{L}, \mathcal{E}^\mathcal{L})$, where $\mathcal{V}^\mathcal{L}$ is the node set and node $i$ represents the $i$-th atom in the ligand. 
In this work, RdKit~\cite{greg_landrum_2022_6798971} is employed to generate a 3D initial conformer of the input ligand.
Each node $v^\mathcal{L}_i$ is also associated with an atom coordinate $x^\mathcal{L}_i$ retrieved from the individual ligand structure $\mathcal{P}$ and an atom feature vector $h^\mathcal{L}_i$.
The edge set $\mathcal{E}^\mathcal{L}$ is constructed according to the spatial distances among atoms. More formally, the edge set is defined to be: 
\begin{equation}
     \mathcal{E}^\mathcal{L} = \left\{(i,j): |x^\mathcal{L}_i - x^\mathcal{L}_j|^2< cut^{\mathcal{L}}, \forall i, j \in \mathcal{V}^\mathcal{L}\right\},
\end{equation}
where $cut^{\mathcal{L}}$ is a distance threshold, and each edge $(i,j)\in \mathcal{E}^\mathcal{L}$ is associated with an edge feature vector $e^\mathcal{L}_{ij}$. The node and edge features are obtained by RDKit~\cite{greg_landrum_2022_6798971} in the \sitemodel. And in the \refinemodel, they are achieved by OpenBabel~\cite{OBoyle2011OpenBA} 

\textbf{Protein Atomic Graph.}
The protein atomic graph $\mathcal{G}^{\mathcal{P}}$ is constructed in the same way as the ligand graph.

\textbf{Protein Residue Graph.}
For protein residue Graph $\mathcal{G}^{\mathcal{P}*}=(\mathcal{V}^{\mathcal{P}*}, \mathcal{E}^{\mathcal{P}*})$,
$\mathcal{V}^{\mathcal{P}*}$ is the node set and the node $i$ represents the $i$-th residue in the protein.
Each node $v^{\mathcal{P}*}_i$ is also associated with an $C_\alpha$ coordinate of the $i$-th residue $x^{\mathcal{P}*}_i$ retrieved from the individual protein structure and a residue feature vector $h^{\mathcal{P}*}_i$. 
The edge set $\mathcal{E}^{\mathcal{P}*}$ is constructed according to the spatial distances among atoms. More formally, the edge set is defined to be:
\begin{equation}
      \mathcal{E}^{\mathcal{P}*} = \left\{(i,j): |x^{\mathcal{P}*}_i - x^{\mathcal{P}*}_j|^2< cut^{\mathcal{P}*}, \forall i, j \in \mathcal{V}^{\mathcal{P}*} \right\},
\end{equation}
where $cut^{\mathcal{P}*}$ is a distance threshold, and each edge $(i,j)\in \mathcal{E}^{\mathcal{P}*}$ is associated with an edge feature vector $e^{\mathcal{P}*}_{ij}$. 
The edge features are obtained following~\cite{Corso2022DiffDockDS}.
As for the node features, they are extracted from the protein language model ESM2-3B~\cite{lin2022language} in \sitemodel.
While in \refinemodel, they are obtained following \cite{Strk2022EquiBindGD}.

\section{More Detailed Experimental Settings}
\label{apd:exp_settings}
\subsection{Baselines}
For molecular docking, GDL methods, EquiBind~\cite{Strk2022EquiBindGD}, TANKBind~\cite{Lu2022TANKBindTN}, DiffDock~\cite{Corso2022DiffDockDS}, DeepDock~\cite{MndezLucio2021AGD}, Uni-Mol~\cite{Zhou2023UniMolAU}, and FABind~\cite{Pei2023FABindFA}, and traditional sampling methods, VINA~\cite{Trott2010AutoDockVI}, SMINA~\cite{Koes2013LessonsLI}, and DSDP~\cite{Huang2023DSDPAB} are used as baselines. 
For pocket prediction, DSDP, P2Rank~\cite{Krivk2018P2RankML},  Fpocket~\cite{Guilloux2009FpocketAO}, SiteMap~\cite{halgren2007new, halgren2009identifying}, and DoGSite3~\cite{graef2023binding} are compared.

\subsection{Evaluation Metric} 
For blind docking and site-specific docking, RMSD and centroid distance are used to evaluate different methods, the formal definitions of these two metrics are:
\begin{align}
    & RMSD = \sqrt{\frac{1}{|V|}\sum_{i=1}^{|V|} (x_i^{\mathcal{L}} - \hat{x}_i^{\mathcal{L}'})^2}, \\
    & Centroid = |\frac{1}{|V|}\sum_{i=1}^{|V|}x_i^{\mathcal{L}} - \frac{1}{|V|}\sum_{i=1}^{|V|}\hat{x}_i^{\mathcal{L}'}|,
\end{align}
where $x_i^{\mathcal{L}}$ is the ground truth coordinate of $i$-th ligand atom and $\hat{x}_i^{\mathcal{L}'}$ is the predicted coordinates.
In alignment with previous studies~\cite{Zhu2023DiffBindFRAS, Corso2022DiffDockDS}, for blind docking, the RMSD is directly computed. However, in the case of site-specific docking, the RMSD is calculated utilizing the spyrmsd~\cite{Meli2020spyrmsdSR}.

For pocket prediction, the DCC metric is defined as:
\begin{equation}
    DCC = |\hat{\varsigma} - \frac{1}{|V|}\sum_{i=1}^{|V|}\hat{x}_i^{\mathcal{L}'}|,
\end{equation}
where $\hat{\varsigma}$ is the predicted pocket center.
As for the VCR metric~\cite{Huang2023DSDPAB}, we calculate the cube side length of a cube box centered on the pocket that can cover the whole ligand structure.

\subsection{Training and inference} 
Our models are trained using NVIDIA A100-PCIE-40GB GPUs. Training the \sitemodel on a single GPU takes approximately 2 hours, while the \refinemodel requires about 48 hours on 4 GPUs. To determine the hyperparameters, we performed a grid search, as outlined in Table \ref{apd:tbl_CPLA_hyper} and Table \ref{apd:tbl_biegmn_hyper}.

\subsubsection{\sitemodel}
\label{apdsec:cpla_exp_settings}
\textbf{Basic Settings.} The model was trained employing the Adam optimizer~\cite{Kingma2015AdamAM} with an initial learning rate of $0.0003$ and an $L_{2}$ regularization factor of $10^{-6}$. The learning rate was scaled down by 0.6 if no drop in training loss was observed for 10 consecutive epochs.
The number of training epochs was set to 20 with an early stopping rule of 10 epochs if no improvement in the validation performance was observed. 

\textbf{Candidate Pockets Generation.} For \sitemodel, we consider two methods to generate candidate pockets: DSDP, and P2Rank. These methods were selected over others, such as SiteMap. Initially, we intended to incorporate all available methods to construct the candidate pockets. However, the results were unsatisfactory. This could be attributed to the issue of hard negative samples. CPLA employs contrastive learning, where the quality of hard negative sample selection directly impacts the training performance. In this context, hard negative samples represent highly druggable pockets that are not the target pocket. As illustrated in Fig.~\ref{fig:apd_pocket_num} and Fig.\ref{fig:apd_dcc_total}, the pockets predicted by DSDP and P2rank are highly druggable. In contrast, other methods tend to predict non-druggable pockets. The result in Table.~\ref{apdtbl:pocket_pred_perf_fpocket} demonstrates that introducing FPocket impairs the training quality. Consequently, we opted to solely use DSDP and P2rank.

\textbf{Pocket Augmentation.} Given a candidate pockets set $S = \{\varsigma_1, \varsigma_2, ...\}$, we establish a maximum pocket number, $N_{max}$, to construct negative pockets for data augmentation. 
If $|S|>=N_{max}$, we select the top-$N_{max}$ pockets in the sort of DSDP, P2Rank accordingly.
If $|S|<N_{max}$, we randomly select $(N_{max} - |S|)$ $C_\alpha$ atoms that are more than 20.0 \AA \; from the ligand geometric center to construct negative pocket centers.
This data augmentation is only applied in the training phase.

\textbf{Ligand Conformation Augmentation.} During the CPLA training, we further considered the issue of the native binding mode. As the native binding mode (i.e., the co-crystal structure) of a given molecule is unknown in practical scenarios, we aim to train a pose-robust CPLA model. To achieve this, we adjusted the rotatable bond angles of the co-crystal molecule structure in each epoch during training. Therefore, the molecule poses in each epoch are perturbed and different. 

\textbf{Other Training Object.} We have considered using cross-protein loss for training, where the ground truth pockets and ligands from the same protein-ligand pairs are considered positive samples, and those from different protein-ligand pairs are treated as negative samples. Although this loss has been utilized in previous work for virtual screening~\cite{Gao2023DrugCLIPCP}, it was found to be unsuitable for our model.

 \begin{table*}[ht]
\centering
\caption{The hyperparameter options we searched through for \sitemodel. The final parameters are marked in \textbf{bold}.}
\label{apd:tbl_CPLA_hyper}
\begin{tabular}{c|c}
\toprule
Parameter & Search Sapce \\
\midrule
Number of layers & 2, \textbf{3}, 4 \\
Batch Size & 8, \textbf{16}, 32, 64, 128 \\
Dropout & \textbf{0.1} \\
Learning rate & 0.003, 0.001, \textbf{0.0003}, 0.0001 \\
Max pocket number for training & Null, 16, \textbf{32}, 64, 128 \\ 
Pocket used for training & \textbf{[DSDP, P2Rank]} \\
Training loss & \textbf{Intra-protein}, Cross-protein \\
ESM2-3B embedding & \textbf{True}, False \\
AFP hidden dimension & 64, \textbf{128}, 256 \\ 
GVP node scalar hidden dimension & 32, \textbf{64}, 128 \\
GVP node vector hidden dimension & 12, \textbf{16}, 32 \\
GVP edge scalar hidden dimension & \textbf{32}, 64, 128 \\
GVP edge vector hidden dimension & \textbf{12}, 16, 32 \\
\bottomrule
\end{tabular}
\end{table*}

\subsubsection{\refinemodel}
\label{apdsec:biegmn_exp_settings}
\textbf{Basic Settings.} The Adam optimizer~\cite{Kingma2015AdamAM}, characterized by an initial learning rate of $10^{-3}$ and an $L_{2}$ regularization factor of $10^{-6}$, is employed for training \refinemodel. The learning rate was scaled down by 0.6 if no drop in training loss was observed for 10 consecutive epochs. The number of training epochs was set to 1000 with an early stopping rule of 40 epochs if no improvement in the validation performance was observed. 

\textbf{Training Object.} The loss function can be written as:
\begin{equation}
     L =
     L_{inter} +
     \lambda_1 L_{intra} +
     \lambda_2 L_{vdw} +
     \lambda_3 L_{bound}.
\end{equation}
As introduced before, the inter-distance map loss $L_{inter}$ is responsible for the RMSD accuracy. Other three items, namely intra-distance map loss $L_{intra}$, vdW constraint loss $L_{vdw}$, and bound matrix constraint loss $L_{bound}$ are employed for physical validity.

The two distance map losses can be formally expressed as:
\begin{align}
     L_{inter} = \sum_{i \in \mathcal{V}_\mathcal{L}}\sum_{j \in \mathcal{V}_\mathcal{P}} ||d_{ij}^{pred} - d_{ij}^{gt}||, \; 
     L_{intra} = \sum_{i \in \mathcal{V}_\mathcal{L}}\sum_{j \in \mathcal{V}_\mathcal{L}} ||d_{ij}^{pred} - d_{ij}^{gt}||,
\end{align}
where predicted distance $d_{ij}^{pred}=||x_i^{pred} - x_j^{pred}||$ and ground-truth distance $d_{ij}^{gt}=||x_i^{gt} - x_j^{gt}||$  between node $i$ and $j$ are calculated based on node coordinates.

The other two physics-informed losses can be formally expressed as:
\begin{align}
     L_{vdw} & = \sum_{i \in \mathcal{V}_\mathcal{L}}\sum_{j \in \mathcal{V}_\mathcal{P}} max(d_{ij}^{vdw} - d_{ij}^{pred}, 0 ), \\
     L_{bound} & = \sum_{i \in \mathcal{V}_\mathcal{L}}\sum_{j \in \mathcal{V}_\mathcal{L}} max(d_{ij}^{bd,low} - d_{ij}^{pred}, 0) + max(d_{ij}^{pred} - d_{ij}^{bd,up}, 0),
\end{align}
where the vdW distance $d_{ij}^{vdw}=0.75 (r_i^{vdw} + r_j^{vdw}) $ is calculated based on node van der Waals radii $r^{vdw}$. As for the lower bound distance $d_{ij}^{bd,low}$ and upper bound distance $d_{ij}^{bd,up}$, they are determined based on the bound matrix generated by RDKit~\cite{landrum2013rdkit} following~\cite{Buttenschoen2023PoseBustersAD}.

\begin{figure}[ht]
 \centering
  \includegraphics[width=0.7\linewidth]{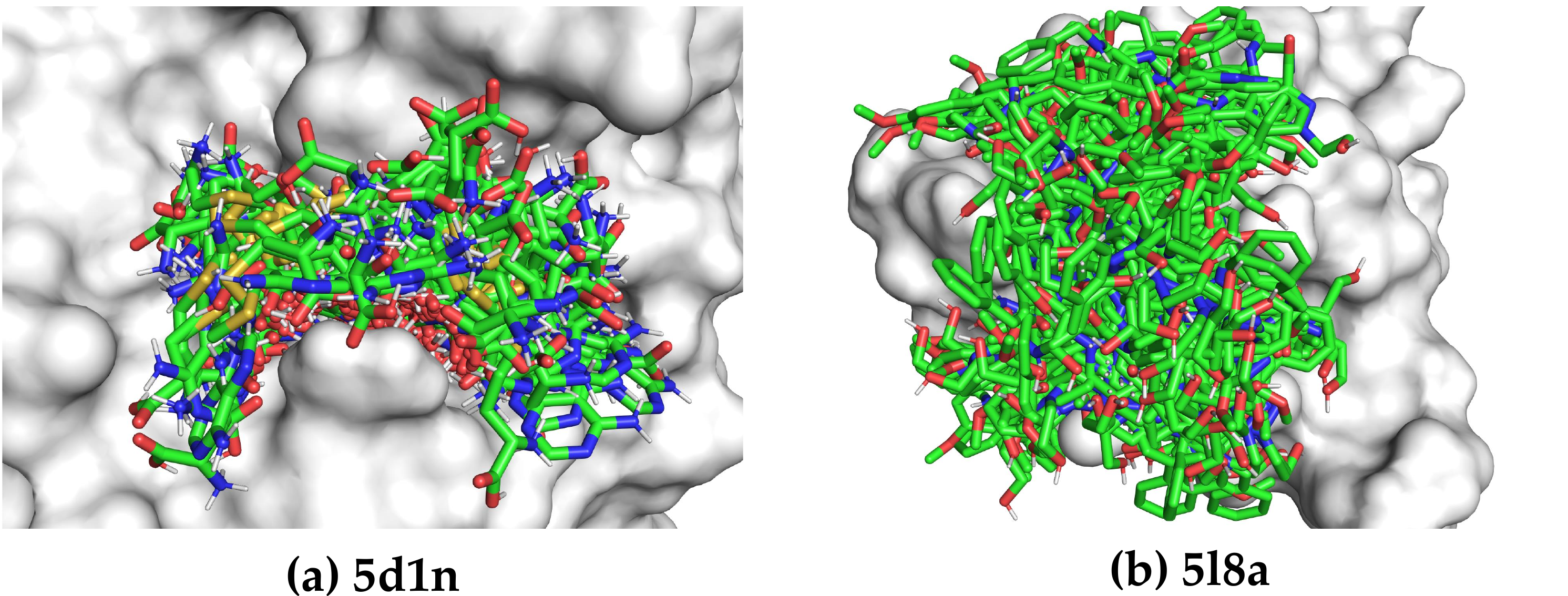}
  \caption{
  Initial pose augmentation. During the initial pose augmentation phase of training the \refinemodel, we randomly select one pose from all sampled poses for each epoch. This selection strategy ensures that the training initial poses can comprehensively cover the entire pocket.}
  \label{fig:apd_pose_aug}
 \end{figure}

\textbf{Initial Poses Augmentation.} In the training phase of the \refinemodel, initial pose augmentation is employed. The initial poses utilized for training are sampled based on the ground truth pocket. An adaptive box is defined through a two-step process: (1) the minimum and maximum of the x, y, and z coordinates of the ligand are extended by 4 \AA \; each; (2) if the box size is less than 22.5 \AA \; after the first step, it is further extended to 22.5 \AA. During the inference phase, however, the box size is fixed at 30.0 \AA, deviating from the adaptive strategy employed during training.
For each epoch during training, a pose is randomly selected. This pose augmentation strategy significantly amplifies the diversity of the input. As depicted in Fig.\ref{fig:apd_pose_aug}, the sampled poses can nearly encompass the entire pocket cavity. 

\textbf{Recycling.} During both training and inferencing, the recycling strategy is adopted. For training, we randomly recycle the iterative refinement process 1-3 times, and only the last cycle is used to compute the gradient.
For inferencing, the recycle number is fixed to 4.

 \begin{table*}[ht]
\centering
\caption{The hyperparameter options we searched through for \refinemodel. The final parameters are marked in \textbf{bold}. }
\label{apd:tbl_biegmn_hyper}
\begin{tabular}{c|c}
\toprule
Parameter & Search Sapce \\
\midrule
Recycle & \textbf{True}, False \\
Hidden dimension & 32, \textbf{64}, 96, 128 \\ 
Number of layers for each level & 4, 6, \textbf{8}, 10 \\
Batch Size & 8, 16, \textbf{32}, 64 \\
Dropout & \textbf{0.1} \\
Learning rate & \textbf{0.001} \\
Initial pose augmentation & \textbf{True}, False \\
Pose sampling box size & Adaptive, \textbf{30.0 \AA}  \\
\sitemodel pockets used for sampling & Top-1, \textbf{Top-2}, Top-3, All \\
Protein structure level & Atom level, Residue level, \textbf{Bi-level} \\
ESM2-3B embedding for residue level & True, \textbf{False} \\
\bottomrule
\end{tabular}
\end{table*}

\subsection{Ablation Studies Settings}
\label{apdsec:ablation_settings}
\textbf{w/o \sitemodel}: pockets predicted by DSDP are employed to perform the following predictions.

\textbf{w/o \refinemodel}: the sampled structures are directly employed as final structures to calculate metrics.

\textbf{w/o Residue Level}: the residue level is removed from \refinemodel.

\textbf{w/o Atom Level}: the atom level is removed from \refinemodel.

\section{More Experimental Results}
\label{apd:add_result}
\subsection{Binding Pocket Prediction}
\subsubsection{Overall Performance on PDBbind}
\begin{figure}[ht]
\centering
  \includegraphics[width=1.0\linewidth]{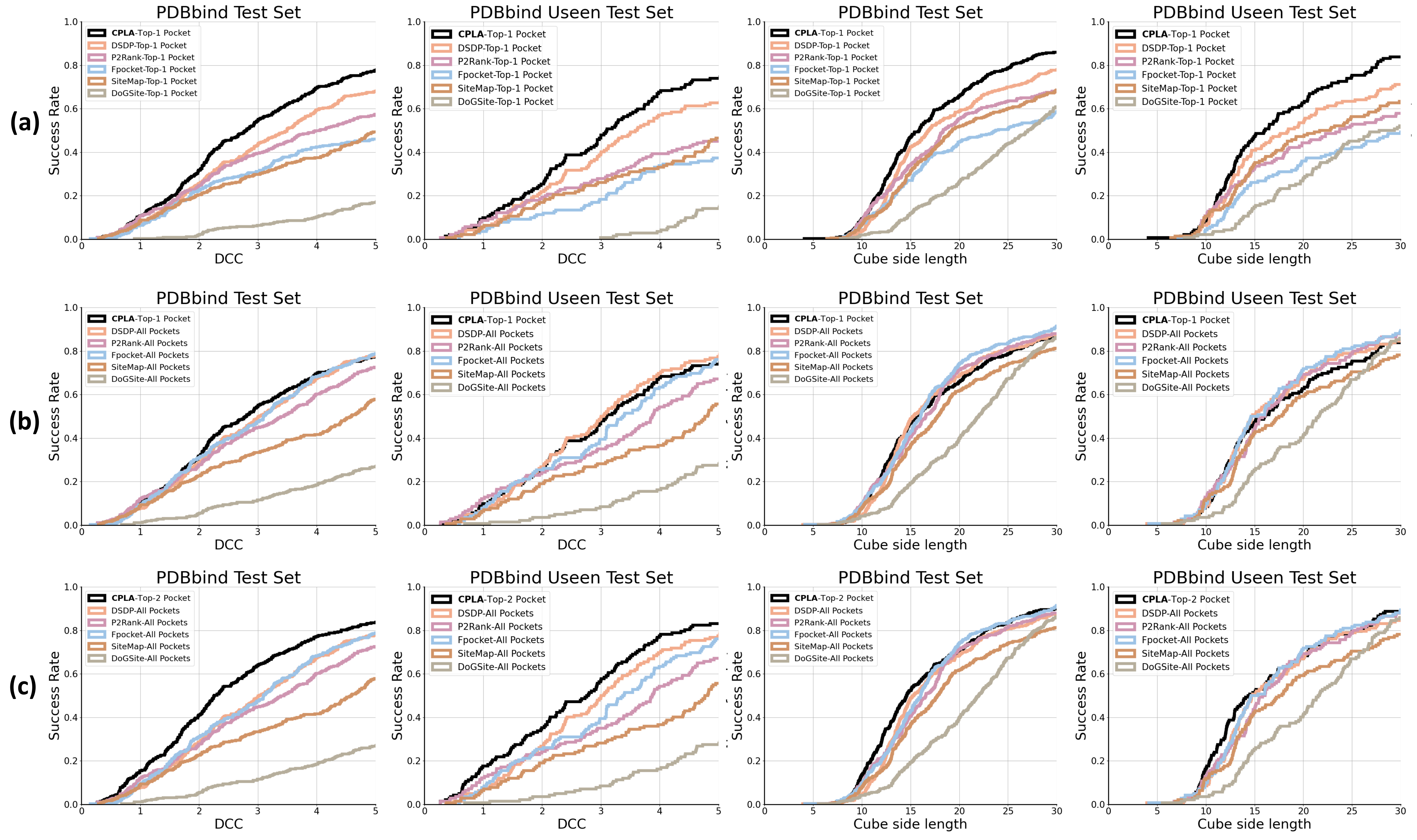}
  \caption{
  Performance of binding pocket prediction models on PDBbind dataset.
  (a) Comparison between top-1 pockets predicted by \sitemodel and top-1 pockets predicted by other methods.
  (b) Comparison between top-1 pockets predicted by \sitemodel and best pockets among all pockets predicted by other methods.
  (c) Comparison between top-2 pockets predicted by \sitemodel and best pockets among all pockets predicted by other methods.
  }
  \label{fig:apd_pocket_pred}
 \end{figure}

In addition to the overall performance presented in Fig.\ref{fig:further_analysis}, we offer a more detailed analysis in Fig.\ref{fig:apd_pocket_pred}. As can be discerned from the figure, the top-1 pockets predicted by \sitemodel significantly outperform those predicted by other baseline methods.
Furthermore, when considering the top-2 pockets, the accuracy of pocket prediction is on par with the cumulative performance of all pockets predicted by other methods.

\subsubsection{Results of Different Candidate Pockets}
In the current framework, only DSDP and P2Rank are selected to generate candidate pockets. The motivation and analysis for this operation have been discussed before. To support this selection, we further present the experimental results of employing different candidate pockets to train \sitemodel in Table.~\ref{apdtbl:pocket_pred_perf_fpocket}. These results indicate that only selecting DSDP and P2Rank yields to best performance.

\begin{table}[ht]
\centering
\caption{Performance of employing different candidate pockets to train \sitemodel}
\label{apdtbl:pocket_pred_perf_fpocket}
\begin{adjustbox}{width=0.5\textwidth}
\begin{tabular}{l|c}
\toprule
\textbf{Pockets} &
\textbf{\% of DCC < 4 \AA }\\
\midrule
DSDP & 64.46 \\
P2Rank & 55.37 \\
DSDP + P2Rank & 69.97 \\
DSDP + P2Rank + Fpocket & 65.84 \\
\bottomrule
\end{tabular}
\end{adjustbox}
\end{table}

\subsubsection{Influence of Ligand Conformations}
When training \sitemodel, we employ a conformation augmentation strategy to train a pose-robust CPLA model.
The provided Table.~\ref{apdtbl:pocket_pred_perf_init_conf} illustrates \sitemodel's performance when presented with both a co-crystal ligand structure and an RDKit-generated ligand structure, showcasing the model's resilience to ligand poses and the effectiveness of our strategy.

\begin{table}[ht]
\centering
\caption{Influence of ligand conformations on \sitemodel}
\label{apdtbl:pocket_pred_perf_init_conf}
\begin{adjustbox}{width=0.4\textwidth}
\begin{tabular}{l|c}
\toprule
\textbf{Input ligand pose} &
\textbf{\% of DCC < 4 \AA }\\
\midrule
Co-crystal & 70.25 \\
RDKit-generated & 69.97 \\
\bottomrule
\end{tabular}
\end{adjustbox}
\end{table}

\subsubsection{Comparison with FABind}
Previous pocket prediction methods, such as DSDP and P2RANK, are ligand-independent. Their goal is to predict all possible binding sites. However, in molecular docking, the goal is to predict targeted binding sites. There are now methods that, like \sitemodel, are ligand-dependent, such as FABind. To further demonstrate the effectiveness of \sitemodel, a comparison is conducted between FABind and \sitemodel as shown in Table.~\ref{apdtbl:comp_fabind}. Our model achieves a significant advantage.

\begin{table}[ht]
\centering
\caption{Comparison with FABind}
\label{apdtbl:comp_fabind}
\begin{adjustbox}{width=0.6\textwidth}
\begin{tabular}{l|c|c}
\toprule
\textbf{Methods} &
\textbf{\% of DCC < 3.0 \AA } & 
\textbf{\% of DCC < 4.0 \AA } \\
\midrule
FABind & 42.7 & 56.5 \\
\sitemodel Top-1 & 54.8 & 70.0 \\
\bottomrule
\end{tabular}
\end{adjustbox}
\end{table}

\subsection{Blind Docking Performance on PoseBusters}
\begin{figure}[ht]
\centering
  \includegraphics[width=0.8\linewidth]{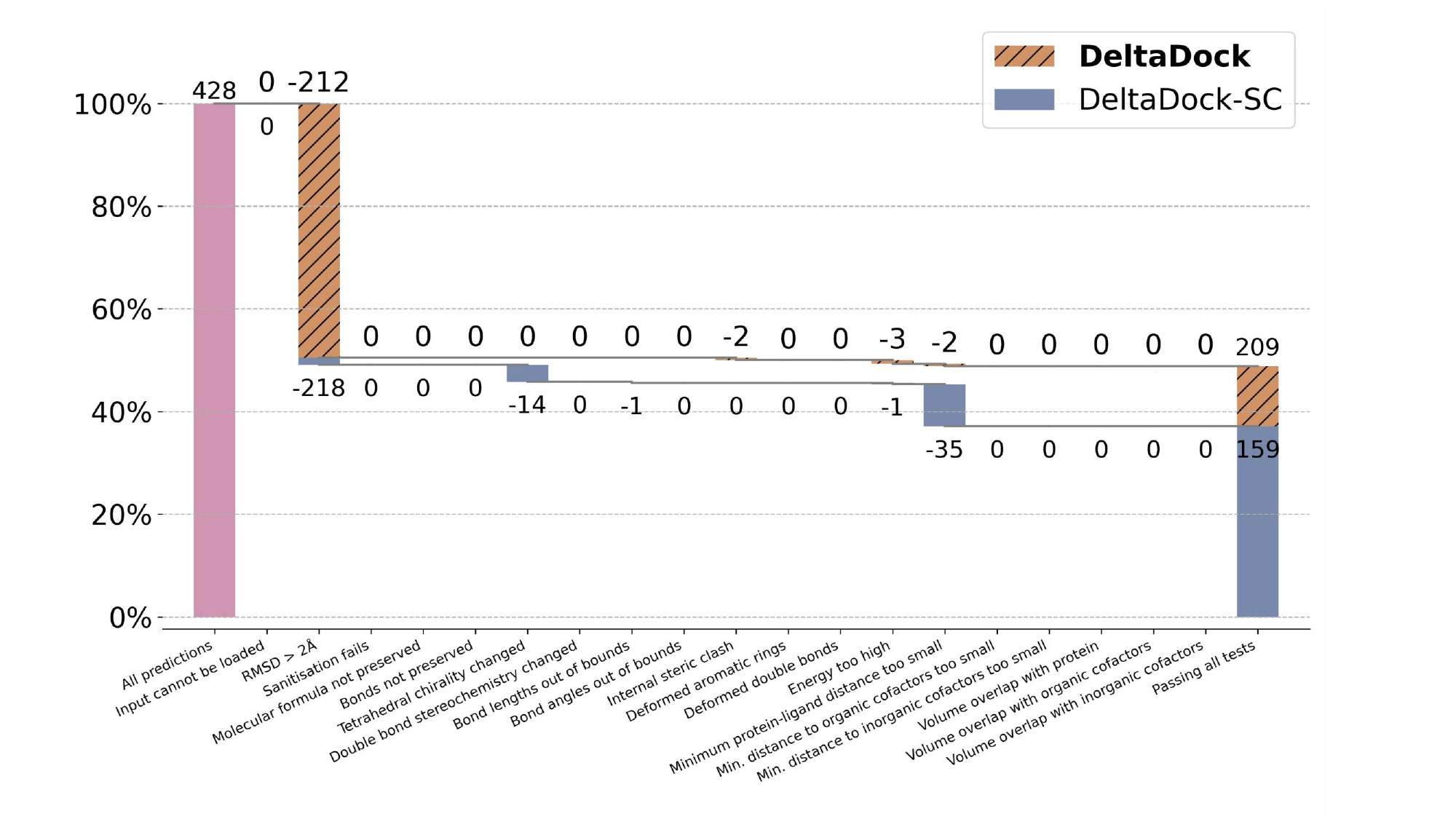}
  \caption{Blind Docking Performance on PoseBusters.}
  \label{fig:apd_blind_dock_posebuster}
 \end{figure}

Due to the space limitation, only site-specific docking performance on PoseBusters has been presented before. In Fig.~\ref{fig:apd_blind_dock_posebuster}, we provide the blind docking performance on PoseBusters. We observed that \modelname achieves a docking success rate of 48.8\% even when considering the physical validity.

\subsection{Detailed Ablation Studies}
\label{apdsec:full_ablation}
Comprehensive ablation experiments were performed within two distinct contexts: blind docking utilizing the PDBbind dataset to assess the impact on RMSD metrics, and site-specific docking employing the PoseBusters dataset to evaluate the influence on the physical plausibility of the predicted binding poses.

\subsubsection{Ablation Studies On PDBbind}
Table.\ref{apdtbl:blind_docking_test_set} presents more detailed ablation studies on PDBbind, including the removal of recycling, training loss components, structure correction, and structure sampling initialization.
From the table, we observe that: (1) each component contributes to the good RMSD performance of our \modelname. (2) The training loss items and structure correction step employed for physical validity tend to decrease the RMSD performance. 
(3) The structure sampling algorithm used for initialization is especially important for good RMSD performance. 
\textbf{(4) When we train \modelname like previous docking methods, removing the loss items and structure correction step for physical plausibility, \modelname still achieves a competitive performance and outperforms all other GDL methods significantly on the test unseen set even without the using of structure sampling algorithm.}
These results demonstrate the effectiveness of \modelname.

 \begin{table*}[ht]
 \centering
 \caption{Blind docking performance on the PDBbind dataset.}
 \label{apdtbl:blind_docking_test_set}
\begin{adjustbox}{width=1.0\textwidth}
\begin{threeparttable}
\begin{tabular}{l|cc|cc|cc|cc}
\toprule
\multirow{3}{*}{Method} &
\multicolumn{4}{c|}{Time Split (363)} & 
\multicolumn{4}{c}{Timesplit Unseen (142)}\\
& \multicolumn{2}{c}{RMSD \% below} 
& \multicolumn{2}{c|}{Centroid \% below}  
& \multicolumn{2}{c}{RMSD \% below} 
& \multicolumn{2}{c}{Centroid \% below} \\
& 2.0\AA  & 5.0\AA  & 2.0\AA  & 5.0\AA  & 2.0\AA  & 5.0\AA & 2.0\AA & 5.0\AA \\
\midrule
\modelname
& 47.4 & 66.9 & 66.7 & 83.2 & 40.8 & 61.3 & 60.4 & 78.9 \\
\midrule
w/o recycle
& 46.0 & 64.2 & 67.2 & 80.2 & 40.8 & 59.9 & 62.0 & 78.2 \\
\midrule
w/o $L_{vdw}$
& 46.8 & 65.3 & 66.4 & 81.3 & 40.8 & 62.7 & 64.8 & 78.2 \\
w/o $L_{intra}$
& 43.5 & 64.7 & 65.0 & 84.8 & 40.1 & 58.5 & 61.3 & 81.7\\ 
w/o $L_{bound}$
& 42.4 & 66.4 & 66.9 & 82.1 & 35.9 & 61.3 & 63.4 & 79.6 \\
\midrule
w/o torsion alignment
& 47.9 & 68.0 & 69.1 & 82.9 & 41.5 & 62.0 & 62.7 & 78.9 \\
w/o energy minimization
& 46.8 & 67.8 & 70.0 & 83.2 & 40.1 & 60.6 & 65.5 & 78.8 \\
w/o structure sampling$^a$ 
& 16.0 & 53.4 & 53.2 & 80.4 & 19.0 & 51.4 & 52.1 & 73.9\\
w/o structure sampling, and structure correction 
& 19.8 & 55.6 & 56.2 & 82.4 & 21.1 & 52.8 & 52.1 & 77.5\\
\midrule
w/o $L_{bound}$, $L_{intra}$, $L_{vdw}$, structure correction, structure sampling
& 30.0 & 63.8 & 65.3 & 82.9 & 28.2 & 53.5 & 57.7 & 78.2 \\

\bottomrule
\end{tabular}
\begin{tablenotes}
 \item[a] No structure sampling means we directly put the RDKit-generated ligand structure at the center of the protein as the initial structure.
 \end{tablenotes}
\end{threeparttable}
\end{adjustbox}

\end{table*}

\subsubsection{Ablation Studies On PoseBusters}
\begin{figure}[ht]
\centering
  \includegraphics[width=0.95\linewidth]{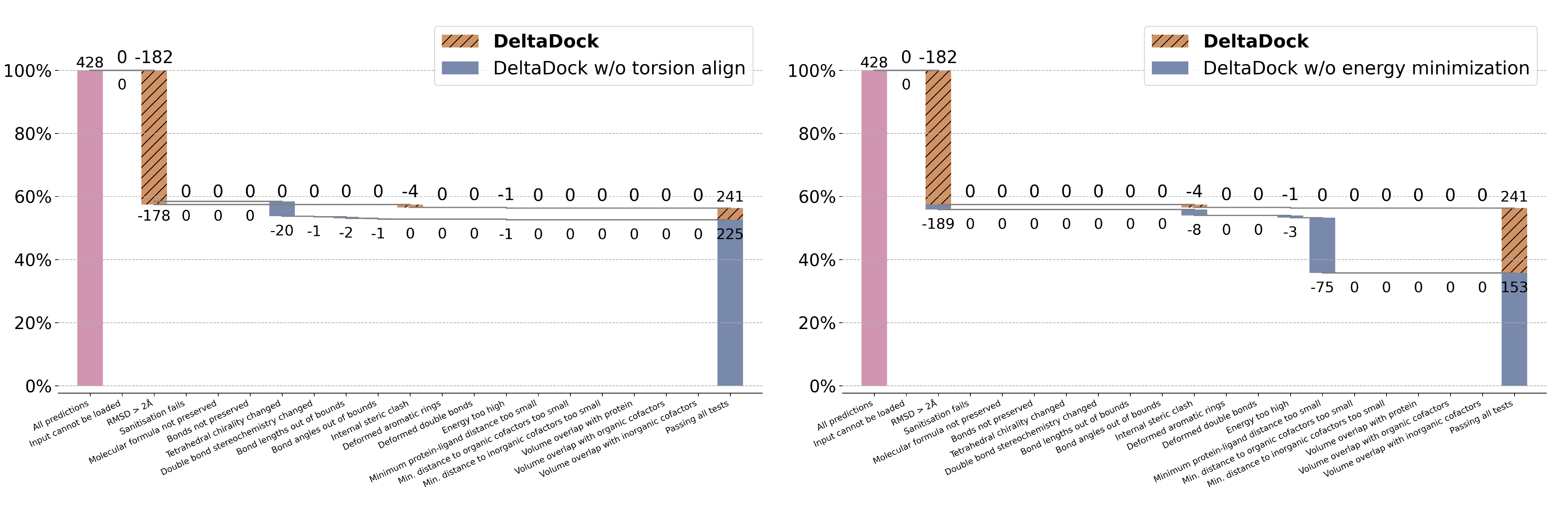}
  \caption{Site-specific docking performance on the PoseBusters dataset.}
  \label{fig:apd_ablation_study_pb_sc}
 \end{figure}

\begin{figure}[ht]
\centering
  \includegraphics[width=0.95\linewidth]{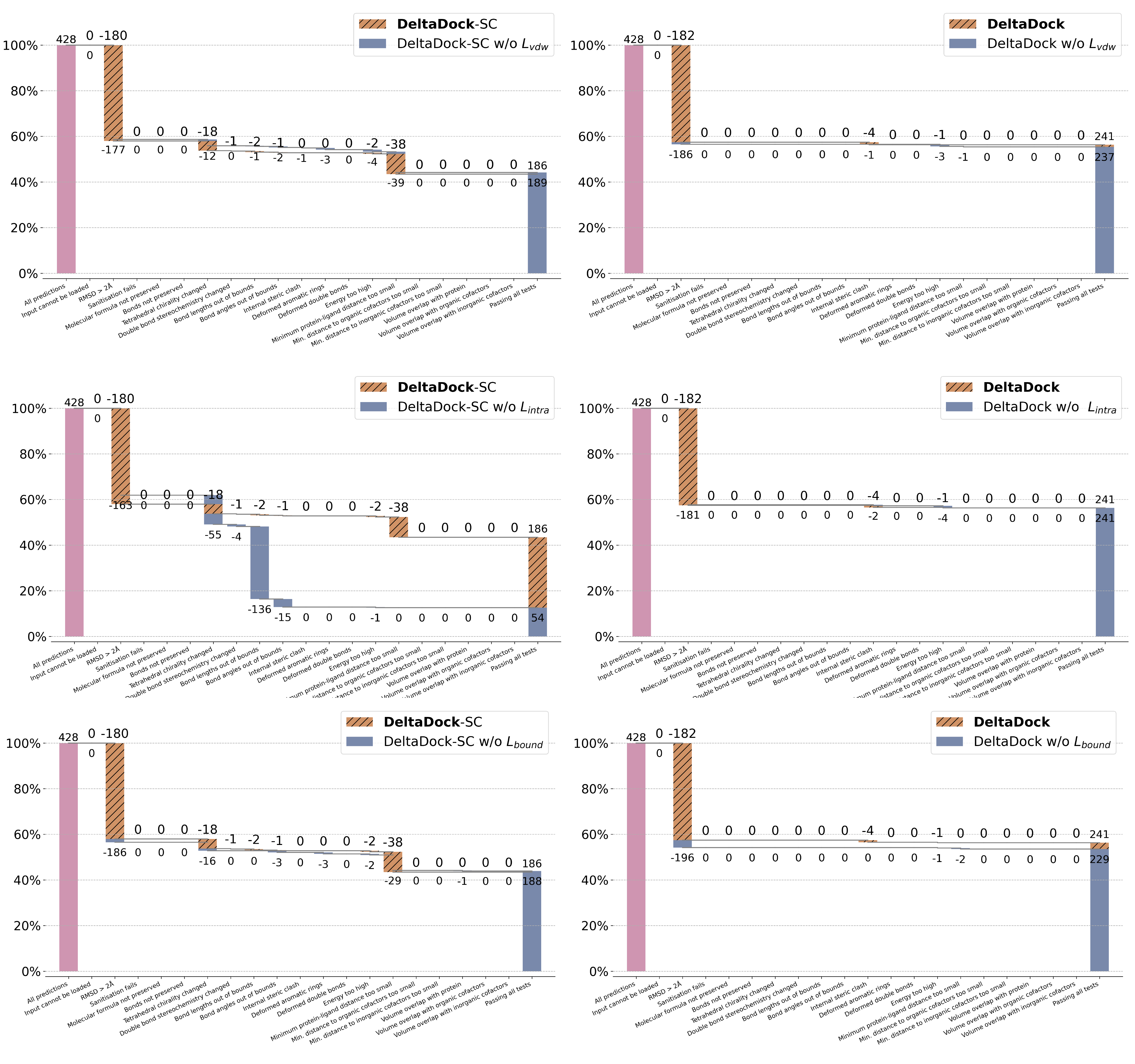}
  \caption{Site-specific docking performance on the PoseBusters dataset.}
  \label{fig:apd_ablation_study_pb_loss}
 \end{figure}

Fig.~\ref{fig:apd_ablation_study_pb_sc} and Fig.~\ref{fig:apd_ablation_study_pb_loss} present ablation studies on PoseBusters to explore the effect of physics-informed training items and structure correction step.
From the figures, we can see that: (1) the physics-informed training items and structure correction step contribute to the good physical validity of \modelname.
(2) Among the physics-informed training items, $L_{intra}$ is especially important for the GDL model to predict valid structures without post-processing.
These results demonstrate the effectiveness of \modelname.

\section{Broader Impacts and Limitations}
\subsection{Broader Impacts}
\label{apdsec:broader_impacts}
The development and maintenance of computational infrastructure for AI-assisted molecular docking represent a significant allocation of resources. Inefficient allocation or underutilization of these resources can potentially result in resource wastage. 

\subsection{Limitations}
\label{apdsec:limitation}
One disappointing limitation is the reliance on external tools, such as SMINA for post-processing and the structure sampling algorithm for structure initialization. Although \modelname still achieves the best performance among GDL methods on the test unseen time split set without these tools, the overall performance degrades. Indeed, due to the limited training data, it's quite difficult to accomplish accurate, efficient, and physically reliable docking without any external tools.
In the future, we will try to overcome this limitation by exploring pre-training strategies on large-scale datasets generated by docking methods or recently developed AlphaFold3~\cite{Abramson2024AccurateSP}.

\end{document}